%% file: ms.tex
\documentclass[12pt]{article}

\usepackage{graphicx}
\usepackage{setspace}
\usepackage[textwidth=5.5cm]{geometry}
\usepackage{adjustbox}
\usepackage{enumitem}
\usepackage{rotating}
\usepackage{amsmath, amsthm, amssymb,bm}
\usepackage{lscape}
\usepackage{empheq}
\usepackage{multirow}
\usepackage{fancyhdr}
\usepackage{booktabs}
\usepackage[usenames,dvipsnames]{color}
\usepackage[pdftex,colorlinks=true]{hyperref}
\usepackage{slashbox}
\usepackage[hang,bf]{caption}
\usepackage{colortbl}
\usepackage[pdftex,colorlinks=true]{hyperref}
\usepackage[latin1]{inputenc}
\usepackage{longtable}
\usepackage{lastpage}
\usepackage{dcolumn}
\usepackage{tabularx}
\usepackage{tabulary}
\usepackage{ifpdf}
\usepackage{float}
\usepackage{color}
\hypersetup{
colorlinks,
citecolor=blue,
linkcolor=red,
urlcolor=Violet}
\usepackage{epstopdf}
\DeclareGraphicsExtensions{.eps}
\usepackage[round]{natbib}

\usepackage{algorithm}
\usepackage{algpseudocode}

\usepackage{geometry}
\theoremstyle{plain}
\newtheorem{prop}{Proposition}[section]

\newtheorem{proposition}[prop]{Proposition}
\newtheorem{lemma}[prop]{Lemma}
\newtheorem{corollary}[prop]{Corollary}

\theoremstyle{definition}

\theoremstyle{remark}
\allowdisplaybreaks

\title{Focused econometric estimation for noisy and small datasets: A Bayesian Minimum Expected Loss estimator approach}

\author{Andr\'es Ram\'irez-Hassan\thanks{Department of Economics, School of Economics and Finance, Universidad EAFIT, Medell\'in, Colombia. email: aramir21@eafit.edu.co}
\thanks{Department of Econometrics and Business Statistics, Monash University, Melbourne, Australia. email: andres.ramirezhassan@monash.edu.}
\and
 Manuel Correa-Giraldo\thanks{Department of Economics, School of Economics and Finance, Universidad EAFIT, Medell\'in, Colombia. email: mcorre33@eafit.edu.co}}

\date{\today}

\begin{document}
	\maketitle \thispagestyle{empty}
	
	\begin{abstract}
		\setlength{\baselineskip}{10pt}
		\vspace{0.2in}
		
		\noindent Central to many inferential situations is the estimation of rational functions of parameters. The mainstream in statistics and econometrics estimates these quantities based on the \textit{plug-in} approach without consideration of the main objective of the inferential situation. We propose the Bayesian Minimum Expected Loss (MELO) approach focusing explicitly on the function of interest, and calculating its frequentist variability. Asymptotic properties of the MELO estimator are similar to the \textit{plug-in} approach. Nevertheless, simulation exercises show that our proposal is better in situations characterized by small sample sizes and noisy models. In addition, we observe in the applications that our approach gives lower standard errors than frequently used alternatives when datasets are not very informative.

		\vspace{0.2in} \noindent
		
	\end{abstract}
	\newpage
	\setcounter{page}{1}
	\pagenumbering{arabic}

	\section{Introduction}\label{sec:intro}
	
	Central to many statistical and econometric inferential situations is the estimation of rational functions of the parameters.
	These functions might be elasticities, forecasts, impulse responses, marginal effects, odds ratios, optimal quantities, or structural parameters, among others.
	The mainstream in statistics and econometrics estimates these quantities based on a {\it{plug-in}} approach, where parameter estimates are just plugged in to the objective expressions without consideration of the main objective of the inferential situation.
	The popularity of this approach is based on the asymptotic properties of the delta method.
	However, this approach suffers from shortcomings, such as infinite moments and unbounded risks, when based on common considerations such as Gaussian likelihoods, and quadratic loss functions.\\
	
	Models have a purpose.
	So, we should optimally design model's estimation frameworks for their purposes.
	This article is concerned with the estimation of rational functions of parameters, but deviates from the mainstream in that we focus the estimation process directly on the quantities of interest.
	This idea has been used by \cite{Claeskens2003} and \cite{Hansen2005} for model selection, and \cite{DiTraglia2016} for selecting moment conditions in the generalized method of moments (GMM).\\
	
	We extend the idea of the Bayesian Minimum Expected Loss (MELO) approach introduced by \cite{Zellner1978}, whose main theoretical developments and applications were confined to structural econometric models.
	We introduce it for problems where the main concerns of the inference are rational functions of the parameters.
	In particular, we follow a decision theoretic framework where the posterior expected value of a generalized quadratic loss function that depends explicitly on the function of interest is minimized.\\
	
	The Bayesian Minimum Expected Loss approach gives point estimates, so the obvious and common answer to define their degree of variability would be to use the posterior density of the function of interest. This is a right answer, if the prior distribution where based on genuine past experience \citep{berger06,Efron2015}. However, we use diffuse and vague priors, and see the MELO as an estimator, then we follow the idea of \cite{Efron2012,Efron2015}, who proposes to estimate the frequentist variability of the Bayesian estimates. In addition, this approach avoids sensitivity analysis of the choice of prior or hierarchical prior structures, and as a consequence their extra computational burden.\\
	
	We find that the asymptotic properties of the MELO estimators are similar to those of the \textit{plug-in} approach. However, simulation exercises suggest that our proposal obtains better outcomes than competing alternatives; especially in settings characterized by noisy models and small sample sizes.
	In addition, we apply our proposal to real datasets, finding that MELO is more efficient than other alternatives.\\ 
	
	The MELO approach has its foundation in statistical decision theory \citep{Wald1945,Wald1947}, which initially was advocated in econometrics by \cite{Marschak1960} and \cite{Dreze1974}.
	It was introduced in econometrics by \cite{Zellner1978}, who analyzed reciprocals and ratios of parameters, and structural parameters in econometric models.
	He showed for these cases that the MELO estimator has, at least, finite first and second moments, and as a consequence finite risk with respect to a generalized quadratic loss function.
	On the other hand, common estimators like indirect least squares (ILS), two stage least squares (2SLS), limited information maximum likelihood (LIML), three stage least squares (3SLS) and full information maximum likelihood (FIML) have infinite moments and infinite risks using quadratic loss functions.
	Further, \cite{Zellner1979} approximate the small sample moments and risk functions of the MELO estimators, and compared them with other estimators.
	\cite{Zellner1980} found that coefficient estimates of structural parameters using MELO are matrix weighted averages of direct least squares (DLS) and 2SLS.
	\cite{Park1982} showed through simulation exercises that for structural parameters, the MELO estimates have more bias than 2SLS.
	However, MELO outperforms 2SLS in criteria like mean squared error (MSE) and mean absolute error (MAE).
	\cite{Swamy1983} analyzed the requirements of prior distributions for reduced form parameters associated with the MELO estimator in undersized sample conditions, that is, situations where the number of exogenous variables in simultaneous equations models exceeds the sample size.
	They found that the conditions for existence of the FIML estimator are more demanding than the conditions to obtain the MELO.
	\cite{Diebold1997} used the MELO approach to do an interesting application related to the response of agricultural supply to movements in expected price.
	They argued that the large variability of previous estimates associated with this phenomenon is due to the infinite moments and multimodal distributions of common frequentist estimators.
	In contrast, the MELO estimator has at least finite first and second moments; however, it also may exhibit multimodal distributions.
	Finally, \cite{Zellner1998} introduces the Bayesian Method of Moments, and related it to the MELO, extending the approach to cases where we have only moment conditions for our inferential problem.
	He presents the resuts of simulation exercises that show that Bayesian estimators perform better than popular frequentist estimators.\\
	
	This paper is structured as follows.
	The next section develops the theoretical framework.
	Section \ref{Sim} exhibits the outcomes of the simulation exercises.
	Section \ref{App} presents the main findings in our applications.
	Finally, we make some concluding remarks.
	
	\section{Theoretic framework}
	Suppose that the main concern of the econometric inference is $\bm{\omega}={\bf{g}}(\bm{\theta}):\bm{\Theta}\subset\mathcal{R}^L\rightarrow \mathcal{R}^K$, $K\leq L$, that is, $\omega=(\omega_1,\omega_2,\ldots\omega_K)^T=(g_1(\bm{\theta}),g_2(\bm{\theta}),\ldots,g_K(\bm{\theta}))^T$, $g_k(\bm{\theta})=\frac{l_k(\bm{\theta})}{m_k(\bm{\theta})}:\mathcal{R}^L\rightarrow \mathcal{R}, k=1,2,\ldots,K$, such that 
	\begin{align*}
	\gamma=\tau(\bm{\theta}):\mathcal{R}^{L}&\longrightarrow \mathcal{R}^L \\
	\bm{\theta} &\longmapsto ({\bf{g}}(\bm{\theta}),\bm{q}(\bm{\theta}))
	\end{align*}
	is a one-to-one continuously differentiable transformation for some nuisance transformation $\bm{\psi}=\bm{q}(\bm{\theta}):\mathcal{R}^{L}\rightarrow \mathcal{R}^{L-K}$.\\
	
	Our view is that such an inferential problem should be directly tackled focusing on the functions of interest.
	So, we propose for this inferential problem the posterior Bayesian action that minimizes the posterior expected value of a generalized quadratic loss function focused on ${\bf{g}}(\bm{\theta})$, that is,  
	\begin{align*}
	\min_{\hat{\omega}\in \mathcal{R}^K}E_{\pi(\bm{\theta}|{\bf{y}})}\left\{\mathcal{L}({\bf{g}}(\bm{\theta}),\hat{\omega})\right\}& =\min_{{\hat{\omega}\in \mathcal{R}^K}}\int_{\Theta}{\left\{\mathcal{L}({\bf{g}}(\bm{\theta}),\hat{\omega})\right\}\pi(\bm{\theta}|{\bf{y}})d\bm{\theta}}
	\end{align*}
	\noindent where $\mathcal{L}({\bf{g}}(\bm{\theta}),\hat{\omega})=({\bf{g}}(\bm{\theta})-\hat{\omega})^T{\bf{Q}}(\bm{\theta})({\bf{g}}(\bm{\theta})-\hat{\omega})$, ${\bf{Q}}(\bm{\theta})=diag\left\{h_k(\bm{\theta})\right\}$, where $h_k(\bm{\theta})$ are case specific weighting functions.\\
	
	\subsection{Assumptions}\label{sec:assumptions}
	Let $Y_1,Y_2,\dots,Y_N$ be iid, each with density $f(y|\bm{\theta})$ with respect to a
	$\sigma$--finite measure $\mu$, where $\bm{\theta}$ is real--valued, and suppose the following regularity conditions hold.
	
	\subsection*{A. Likelihood}
	
	\begin{enumerate}[label=\alph*.]
		\item The parameter space $\bm{\Theta}$ is an open subset of $\mathcal{R}^L$.
		\item The set $A=\left\{y:f(y|\bm{\theta})>0\right\}$ is independent of $\bm{\theta}$.
		\item For every $y\in A$, the density $f(y|\bm{\theta})$ is twice differentiable with respect to $\bm{\theta}$, and the second derivative is continuous in $\bm{\theta}$.
		\item The Fisher information $I(\bm{\theta})=E_{\bm{\theta}}\left[\frac{\partial}{\partial \bm{\theta}}logf(Y|\bm{\theta}) \frac{\partial}{\partial \bm{\theta}^T}logf(Y|\bm{\theta})\right]$ satisfies $0<[I(\bm{\theta})_{ij}]<\infty, i,j=1,2,\dots,L$, where $[A_{ij}]$ denotes element $ij$ of matrix $\bm{A}$.
		\item The integral $\int f(y|\bm{\theta}) d\mu(y)$ can be twice differentiated with respect to $\bm{\theta}$ under the integral sign. This ensures that for all $\bm{\theta}\in\Theta$, $E\left[\frac{\partial}{\partial \bm{\theta}}log f(Y|\bm{\theta})\right]=\bm{0}$ and $E\left[-\frac{\partial^2}{\partial \bm{\theta}\partial{\bm{\theta}}^T}log f(Y|\bm{\theta})\right]=I(\bm{\theta})$. 
		\item For any given $\bm{\theta}_0\in\Theta$, there exists a positive number $c$ and a function $M(y)$ (both of which may depend on $\bm{\theta}_0$) such that $|\partial^2 log f(y|\bm{\theta})/\partial \theta_i\partial\theta_j|\leq M(y)$ for all $y\in A$, $||\bm{\theta}-\bm{\theta}_0||<c$, where $||\cdot||$ is the Euclidean norm, and $E_{\bm{\theta}_0}M(Y)<\infty$.	 
	\end{enumerate}
	
	Under these assumptions is well know that the maximum likelihood estimator satisfies $\sqrt{N}(\hat{\bm{\theta}}-\bm{\theta}_0)\xrightarrow{d} \mathcal{N}\left(0, I(\bm{\theta}_0)^{-1}\right)$, that is, $\hat{\bm{\theta}}$ is consistent for the true value $\bm{\theta}_0$, and asymptotically efficient \citep{Lehmann2003}. Then $\frac{1}{N}[R_N(\bm{\theta})_{ij}]\xrightarrow{p}\bm{0}$ in the second order Taylor series expansion
	\begin{equation}\label{eq:exp}
	l(\bm{\theta})=l(\bm{\theta}_0)+(\bm{\theta}-\bm{\theta}_0)^T\frac{\partial l}{\partial \bm{\theta}}\biggr\rvert_{\bm{\theta}_0} + \frac{1}{2}(\bm{\theta}-\bm{\theta}_0)^T\left[NI(\bm{\theta}_0)+R_N(\bm{\theta})\right](\bm{\theta}-\bm{\theta}_0)
	\end{equation}
	\noindent where $l(\bm{\theta})=log(f(\bm{y}|\bm{\theta}))$ is the log likelihood, $\bm{y}=[y_1,y_2,\dots,y_N].$
	However, Bayesian estimators involve an integral over the whole range of $\bm{\theta}$ values, then it is necessary the following assumption. 
	
	\subsection*{B. Taylor series expansion}
	
	\begin{enumerate}[label=\alph*.]
		\item Given any $\epsilon>0$, there exist $\delta>0$ such that in the expansion \ref{eq:exp}, $$\lim_{N\to\infty}P\left(sup\left\{\left|\frac{1}{N}[R_N(\bm{\theta})_{ij}]\right|:||\bm{\theta}-\bm{\theta}_0||\leq \delta\right\}\geq \epsilon\right)=0.$$	 
	\end{enumerate}
	
	\subsection*{C. Log likelihood bounded contribution}
	
	\begin{enumerate}[label=\alph*.]
		\item For any $\delta>0$, there exist $\epsilon>0$ such that $$\lim_{N\to\infty}P\left(sup\left\{\frac{1}{N}[l(\bm{\theta})-l(\bm{\theta}_0)]:||\bm{\theta}-\bm{\theta}_0||\geq \delta\right\}\leq -\epsilon\right)=1.$$	 
	\end{enumerate}
	This assumption implies that the log likelihood contribution of $\bm{\theta}\notin \mathcal{B}_{\delta}(\bm{\theta}_0)$, where $\mathcal{B}_{\delta}(\bm{\theta}_0)$ is an open ball centered at $\bm{\theta}_0$ with radius $\delta$, is negligible as $N\rightarrow\infty$.
	
	\subsection*{D. Prior density}
	
	\begin{enumerate}[label=\alph*.]
		\item The prior density $\pi(\bm{\theta})$ is continuous and positive for all $\bm{\theta}\in\bm{\Theta}$.
		\item The expectation and second moment of $\bm{\theta}$ under $\pi$ exists, that is, $\int ||\bm{\theta}||^2 \pi(\bm{\theta})d\bm{\theta}<\infty$. 	 
	\end{enumerate}
	
	First assumption implies $\pi(\bm{\theta}_0)>0$, so $\bm{\theta}_0$ is not a priori excluded. The second assumption is required to proof that the minimum expected loss estimator is consistent and asymptotically efficient. We assume that there is a proper prior density function. However, result \ref{eq:main} can be extended to the case $\int\pi(\bm{\theta})d\bm{\theta}=\infty$ when there is $n_0$, such that the posterior density taking information up to this point is a proper density with probability 1, satisfying assumptions \textbf{D}.
	
	\subsection*{E. Objective function}
	
	\begin{enumerate}[label=\alph*.]
		\item $g_k(\bm{\theta})= \frac{l_k(\bm{\theta})}{m_k(\bm{\theta})}:\mathcal{R}^L\rightarrow\mathcal{R}$ is a function with finite and nonzero first order derivative at $\bm{\theta}_0$, continuous first order derivative, and $g_k(\bm{\theta}_0)\neq 0$,  
		such that    $g_k(\bm{\theta})=g_k(\bm{\theta}_0)+(\bm{\theta}-\bm{\theta}_0)^T[\nabla g_k(\bm{\theta}_0)+W_N(\bm{\theta})]$, and $sup\left\{||W_N(\bm{\theta})||:\bm{\theta}\in\bm{\Theta}\right\}<c_1<\infty$, $N\rightarrow\infty$.	 
	\end{enumerate}
	
	\subsection*{F. Weighting functions}
	
	\begin{enumerate}[label=\alph*.]
		\item $h_k(\bm{\theta}):\mathcal{R}^L\rightarrow\mathcal{R}^{++}$ is a function with finite and nonzero first order derivative at $\bm{\theta}_0$, continuous first order derivative, and $h_k(\bm{\theta}_0)\neq 0$, such that    $h_k(\bm{\theta})=h_k(\bm{\theta}_0)+(\bm{\theta}-\bm{\theta}_0)^T[\nabla h_k(\bm{\theta}_0)+V_N(\bm{\theta})]$, and $sup\left\{||V_N(\bm{\theta})||:\bm{\theta}\in\bm{\Theta}\right\}<c_2<\infty$, $N\rightarrow\infty$.	 
	\end{enumerate}
	
	It is well known that assuming \textbf{A} to \textbf{D}, if $\pi^*(\bm{u}|\bm{y})$ is the posterior density of $\bm{u}=\sqrt{N}(\bm{\theta}-\hat{\bm{\theta}})$, then
	\begin{equation}\label{eq:main}
	\int (1+||\bm{u}||^r)\left|\pi^*(\bm{u}|\bm{y})-\phi(\bm{I}(\bm{\theta}_0)^{-1},\bm{u})\right|d\bm{u}\xrightarrow{p}0, \ 0\leq r\leq 2,
	\end{equation}
	
	\noindent where $\phi(B,\bm{x})$ is the density function of a multivariate normal distribution with mean $\bm{0}$ and covariance matrix $B$ \citep{Bickel1969,Lehmann2003}.\\

	
	\begin{proposition}\label{proposition1}
		The posterior Bayesian action, that is, the Minimum Expected Loss estimate, associated with $\mathcal{L}({\bf{g}}(\bm{\theta}),\hat{\omega})$ is 
		\begin{align}\label{TheMost}
		\hat{\omega}^*({\bf{y}})&=\left[E_{\pi(\bm{\theta}|{\bf{y}})}{\bf{Q}}(\bm{\theta})\right]^{-1}E_{\pi(\bm{\theta}|{\bf{y}})}\left[{\bf{Q}}(\bm{\theta}){\bf{g}}(\bm{\theta})\right]\\
		&=\left[\int_{\Theta}{\bf{Q}}(\bm{\theta})\pi(\bm{\theta}|{\bf{y}})d\bm{\theta}\right]^{-1}\left[\int_{\Theta}{\bf{Q}}(\bm{\theta}){\bf{g}}(\bm{\theta})\pi(\bm{\theta}|{\bf{y}})d\bm{\theta}\right]\nonumber
		\end{align}
		
		\noindent where $\pi(\bm{\theta}|\bm{y})=\frac{\pi(\bm{\theta})f(\bm{y}|\bm{\theta})}{\int_{\bm{\Theta}}\pi(\bm{\theta})f(\bm{y}|\bm{\theta})d\bm{\theta}}$.\\
		
		Provided previous assumptions on ${\textbf{Q}}(\bm{\theta})$ and $\bm{g}(\bm{\theta})$, and integration and differentiation can be interchanged. 
	\end{proposition}
	Proof in Appendix \ref{Proof:proposition1}.\\
	
	Observe that our MELO estimate is a weighted average of ${\bf{g}}(\bm{\theta})$, whose weights are given by $\left[\int_{\Theta}{\bf{Q}}(\bm{\theta})\pi(\bm{\theta}|{\bf{y}})d\bm{\theta}\right]^{-1}{\bf{Q}}(\bm{\theta})$.
	These weights implicitly depend on the probability associated with each $\bm{\theta}$ in their parameter space as well as their magnitude.
	When ${\bf{Q}}$ does not depend on $\bm{\theta}$, which implies equal weight to each $\bm{\theta}$, the Minimum Expected Loss estimate is the posterior mean, that is, $\hat{\omega}^*({\bf{y}})=E_{\pi(\bm{\theta}|{\bf{y}})}{\bf{g}}(\bm{\theta})$.\\
	
	A good advantage of the MELO estimates is that they can be easily calculated from the draws of the posterior distributions, $\bm{\theta}_s\sim \pi(\bm{\theta}|{\bf{y}})$, and given $S\rightarrow \infty$, $\frac{1}{S}\sum_{s=1}^S {\bf{Q}}(\bm{\theta}_s)\xrightarrow{p}E_{\pi(\bm{\theta}|{\bf{y}})}{\bf{Q}}(\bm{\theta})$ and $\frac{1}{S}\sum_{s=1}^S {\bf{Q}}(\bm{\theta}_s){\bf{g}}(\bm{\theta}_s)\xrightarrow{p}E_{\pi(\bm{\theta}|{\bf{y}})}\left[{\bf{Q}}(\bm{\theta}){\bf{g}}(\bm{\theta})\right]$ by the law of the large numbers, then
	\begin{equation}\label{TheSMost}
	\hat{\omega}^*_S({\bf{y}})=\left[\frac{1}{S}\sum_{s=1}^S {\bf{Q}}(\bm{\theta}_s)\right]^{-1}\left[\frac{1}{S}\sum_{s=1}^S {\bf{Q}}(\bm{\theta}_s){\bf{g}}(\bm{\theta}_s)\right]
	\end{equation}    
	
	\noindent converges in probability to $\hat{\omega}^*$ by Slutsky's theorem.\\
	
	Observe that $\hat{\omega}^*_{k,S}({\bf{y}})=\sum_{s=1}^Sw_{ks}g_{k}(\bm{\theta}_s)$ where $w_{ks}=\frac{h_k(\bm{\theta}_s)}{\sum_{s=1}^Sh_k(\bm{\theta}_s)}$, that is, the MELO is a weighted average. \cite{Casella1998} show that weighted average estimators may perform better than unweighted average estimators, when evaluated under squared error loss functions. Obviously, this depends on the choice of $w_{ks}$. In particular, as our objective function is rational, there are singularities when $m_k(\bm{\theta})=0$, and as a consequence, $g_k(\bm{\theta})$ is not integrable under the posterior distribution. Therefore, if $\hat{\omega}^*_S({\bf{y}})$ is built such that puts less weight on the draws near singularity points, then the MELO estimator gains stability. In particular, setting $\mathcal{L}(g_k(\bm{\theta}),\hat{\omega}_k)=\epsilon^2_k$, where $\epsilon_k=\hat{\omega}_k m_k(\bm{\theta})-l_k(\bm{\theta})$ is an estimation error, such that $\hat{\omega}_k=g_k(\bm{\theta})$ implies $\epsilon_k=0$, then $h_k(\bm{\theta})=m_k(\bm{\theta})^2$.\\
	
	From a frequentist perspective, there are many situations when the posterior Bayesian actions are equal to the Bayes rules, that is, the estimators that minimize the Bayes risk, $r(\pi_{\bm{\theta}},\hat{\omega})=\int_{\Theta}\int_{Y}\mathcal{L}({\bf{g}}(\bm{\theta}),\hat{\omega})f_Y({\bf{y}}|\bm{\theta})dy\ \pi(\bm{\theta})d\bm{\theta}$.
	However, there are situations where the Bayes risk is infinite, for instance using improper priors in conjunction general quadratic loss functions, and as a consequence, the Bayes rule does not exist.
	Nevertheless, it is still possible to obtain the posterior Bayesian action.\\%
	
	\begin{proposition}\label{propositionNew}
		If \textbf{A} to \textbf{F} hold, and  
		\begin{align*}
		\hat{\mathbf{\omega}}^*_k&=\frac{E_{\pi(\bm{\theta}|{\bf{y}})}[{g_k}(\bm{\theta}) h_k(\bm{\theta})]}{E_{\pi(\bm{\theta}|{\bf{y}})}[h_k(\bm{\theta})]}\\
		&=\mathop{\int}_{\Theta}{g}_k(\bm{\theta})\frac{h_k(\bm{\theta})}{\int_{\Theta}h_k(\bm{\theta})\pi(\bm{\theta}|{\textbf{y}})d\bm{\theta}}\pi(\bm{\theta}|{\textbf{y}})d\bm{\theta},
		\end{align*}
		
		\noindent then,
		$$\sqrt{N}(\hat{\mathbf{\omega}}^*_k-g_k(\bm{\theta}_0))\xrightarrow{d}\mathcal{N}(\bm{0},\nabla g_k(\bm{\theta}_0)^T[I(\bm{\theta}_0)]^{-1}g_k(\bm{\theta}_0)),$$
		so that $\hat{\mathbf{\omega}}^*_k$ is consistent and asymptotically efficient.   	 
	\end{proposition}

	Proof: See Appendix \ref{Proof:propositionNew}.\\
	
	Proposition \ref{propositionNew} establishes that asymptotically, the MELO has similar characteristics to the maximum likelihood estimator. However, it seems that in noisy finite samples the MELO has better properties.\\
	
	If uninformative priors are used, as in our exercises, it seems convenient to estimate the frequentist variability of the MELO estimator, as they were not based on genuine past experience \citep{berger06,Efron2012,Efron2015}. In addition, this approach avoids sensitivity analysis of the choice of priors, or hierarchical Bayesian models, both imposing an extra computational burden, at the cost of requiring sufficient statistics.\\
	
	To accomplish this task we have the following result.
	
	\begin{proposition}
		If $\hat{\bm{\theta}}({\bf{y}})\in \mathcal{R}^P$ is a sufficient statistic for $f_Y({\bf{y}}|\bm{\theta})$, then
		\begin{align}\label{FactProposition}
		\hat{\omega}^*({\bf{y}})&=\hat{\omega}^*(\hat{\bm{\theta}}({\bf{y}}))
		\end{align}
		\noindent where $\hat{\omega}^*(\hat{\bm{\theta}}({\bf{y}}))=\left[E_{\pi(\bm{\theta}|\hat{\bm{\theta}}({\bf{y}}))}{\bf{Q}}(\bm{\theta})\right]^{-1}E_{\pi(\bm{\theta}|\hat{\bm{\theta}}({\bf{y}}))}\left[{\bf{Q}}(\bm{\theta}){\bf{g}}(\bm{\theta})\right]$.
		
	\end{proposition}
	Proof: See Appendix \ref{Proof:proposition2}.\\
	
	Equality \ref{FactProposition} shows that our MELO estimate can be obtained from the posterior distribution associated with the data or its sufficient statistic.
	The resulting data reduction helps to estimate the frequentist variability of the MELO.\\
	
	Setting   
	\begin{equation}\label{alpha}
	\alpha_{\hat{\bm{\theta}}({\bf{y}})}(\bm{\theta}) = \nabla_{\hat{\bm{\theta}}({\bf{y}})} log{f(\hat{\bm{\theta}}({\bf{y}})|\bm{\theta})} = \left(\frac{\partial}{\partial \hat{\bm{\theta}}({\bf{y}})_{1}} log{f(\hat{\bm{\theta}}({\bf{y}})|\bm{\theta})},\, \cdots\, , \frac{\partial}{\partial \hat{\bm{\theta}}({\bf{y}})_{P}} log{f(\hat{\bm{\theta}}({\bf{y}})|\bm{\theta})} \right)
	\end{equation}
	
	we have the following useful result.
	
	\begin{lemma}\label{lema1}
		Given ${\bf{Q}}(\bm{\theta})$ and ${\bf{g}}(\bm{\theta})$, the gradient of $\hat{\omega}^*(\hat{\bm{\theta}}({\bf{y}}))$ is
		
		\begin{align}
		\nabla_{\hat{\bm{\theta}}({\bf{y}})} \hat{\omega}^*(\hat{\bm{\theta}}({\bf{y}}))  = & \left\{E_{\pi(\hat{\bm{\theta}}({\bf{y}}))}[{\bf{Q}}(\bm{\theta})]\right\}^{-1} \\ \nonumber
		\times &  \left\{E_{\pi(\hat{\bm{\theta}}({\bf{y}}))}[({\bf{Q}}(\bm{\theta}){\bf{g}}(\bm{\theta}))\otimes\alpha_{\hat{\bm{\theta}}({\bf{y}})}(\bm{\theta})] - \left[E_{\pi(\hat{\bm{\theta}}({\bf{y}}))}[{\bf{Q}}(\bm{\theta})\otimes\alpha_{\hat{\bm{\theta}}({\bf{y}})}(\bm{\theta})] \right] \left[\hat{\omega} \otimes I_{P}\right]\right\} \label{grad} 
		\end{align}

		\noindent where $I_{P}$ is the identity matrix of order $P$, and the operator $\otimes$ denotes the Kronecker product.
	\end{lemma}
	See the proof in Appendix \ref{Proof:lemma1}.\\
	
	\begin{corollary}\label{corollary}
		When ${\bf{Q}}(\bm{\theta})$ and ${\bf{g}}(\bm{\theta})$ are in $\mathcal{R}$, then 
		
		\begin{align}
		\nabla_{\hat{\bm{\theta}}({\bf{y}})} \hat{\omega}^*(\hat{\bm{\theta}}({\bf{y}})) = &\frac{E_{\pi(\bm{\theta}|\hat{\bm{\theta}}({\bf{y}}))}[Q(\bm{\theta}) g(\bm{\theta}) \alpha_{\hat{\bm{\theta}}({\bf{y}})}(\bm{\theta})|\hat{\bm{\theta}}({\bf{y}})]}{E_{\pi(\bm{\theta}|\hat{\bm{\theta}}({\bf{y}}))}[Q(\bm{\theta})|\hat{\bm{\theta}}({\bf{y}})]} \\ \nonumber
		- & \frac{E_{\pi(\bm{\theta}|\hat{\bm{\theta}}({\bf{y}}))}[Q(\bm{\theta}) g(\bm{\theta})|\hat{\bm{\theta}}({\bf{y}})] E_{\pi(\bm{\theta}|\hat{\bm{\theta}}({\bf{y}}))}[Q(\bm{\theta}) \alpha_{\hat{\bm{\theta}}({\bf{y}})}(\bm{\theta})|\hat{\bm{\theta}}({\bf{y}})]}{(E_{\pi(\bm{\theta}|\hat{\bm{\theta}}({\bf{y}}))}[Q(\bm{\theta})|\hat{\bm{\theta}}({\bf{y}})])^2}
		\end{align}
	\end{corollary}
	
	See the proof in Appendix \ref{Proof:corollary1}.\\
	
	Lemma \ref{lema1} allows calculating the frequentist variability of the MELO estimate \ref{TheMost} through the delta method.\\
	
	\begin{proposition}\label{proposition5}
		Setting $\hat{\bm{\theta}}({\bf{y}})\sim(\mu_{\bm{\theta}},\Sigma_{\bm{\theta}})$, the frequentist covariance matrix of $\hat{\omega}^*({\bf{y}})$ is
		\begin{equation}\label{variability}
		Var(\hat{\omega}^*({\bf{y}}))=Var(\hat{\omega}^*(\hat{\bm{\theta}}({\bf{y}})))\approx \nabla_{\hat{\bm{\theta}}}\hat{\omega}^*(\hat{\bm{\theta}}) \Sigma_{\hat{\bm{\theta}}} \nabla_{\hat{\bm{\theta}}}\hat{\omega}^*(\hat{\bm{\theta}})^T
		\end{equation}
		\noindent provided that $N\rightarrow\infty$, $\hat{\bm{\theta}}\xrightarrow{p}\bm{\theta}$.\footnote{This condition is satisfied in all our examples. In addition, delta method extensions where the derivative of the objective function is not continuous at $\bm{\theta}_0$, but the objective function is directionally differentiable at $\bm{\theta}_0$, are developed by \cite{Fang2015}.}
		
	\end{proposition}
	
	See the proof in Appendix \ref{Proof:proposition3}.\\
	
	The setting of our formulation establishes Proposition \ref{proposition1} as an optimal point estimate for functions of parameters. In the case that an analytical solution does not exist, we can use draws of the posterior distributions to obtain the estimates (Equation \ref{TheSMost}). Proposition \ref{proposition5} allows obtaining the frequentist variance of our Bayesian estimate, provided a sufficient statistic.\\
	
	
	%
	\section{Simulation exercises}\label{Sim}
	
	\subsection{Optimal input}
	
	We consider a very simple problem where a firm is interested in finding the level of input ($x$) that maximizes its profit, where the production function is quadratic, that is, $y = \beta_{1} x + \beta_{2} x^{2}$.
	So, the problem is 
	\begin{equation*}
	\max_{x}\Pi(x)= \max_{x} IT(x) - CT(x) = \max_{x}  p(\beta_{1} x + \beta_{2} x^{2}) - CF - w x
	\end{equation*}
	where $p$ is the product's price, $CF$ represents the fixed costs, and $w$ is the input's price.\\
	
	Then the optimal input is given by
	\begin{equation}\label{main}
	x^{Opt} = \frac{1}{2 \beta_{2}} \left[\frac{w}{p} - \beta_{1}\right]
	\end{equation}
	
	This implies that the optimal production and profit are $y^{Opt}=\frac{1}{2\beta_2}\left[\left(\frac{w}{p}\right)^2-\beta_1^2\right]$ and $\Pi^{Opt}=\frac{\beta_1}{2\beta_2}\left[w-\beta_1p\right]-CF$, respectively.\\	
	Suppose that the decision problem is to find the optimal level of input (Equation \ref{main}), that is, ${\bf{g}}(\bm{\theta})=\omega(\beta_1,\beta_2)=x^{Opt}$.
	
	We can exploit the variability between $y_i$ and $x_i$ in the product function, the variability between $x_i$ and $w_i/p_i$ or $y_i$ and $w_i/p_i$ in the optimal input or production functions, or the variability between $\Pi_i$, $w_i$ and $p_i$ in the optimal profit function to obtain estimates of $\beta_1$ and $\beta_2$.
	The choice depends on assumptions regarding the rationality of the firms as well as the availability of the data.\\
	
	We propose to formulate the mean deviation model associated with the production function to obtain the parameter estimates $\beta=\left[\beta_1\ \beta_2\right]'$.
	In particular, $y_i-\bar{y}=\beta_1(x_i-\bar{x})+\beta_2(x_i^2-\overline{x^2})+u_i$, where $\bar{y}=(1/N)\sum_{i=1}^Ny_i$, $\bar{x}=(1/N)\sum_{i=1}^Nx_i$, $\overline{x^2}=(1/N)\sum_{i=1}^Nx^2_i$ and $u_i\sim\mathcal{N}(0,\sigma^2)$, $i=1,2,\ldots,N$.\\
	
	The likelihood function of this model is
	\begin{equation*}
	f(\beta,\sigma|y, X) \propto \sigma^{-N} exp\left\{- \left[vs^{2} + \left(\beta - \hat{\beta}\right)^T X^TX \left(\beta - \hat{\beta}\right)\right]/ 2 \sigma^{2} \right\}
	\end{equation*}
	where $X$ is the design matrix, $q=dim\{\beta\}$, $v=N-q$, $\hat{\beta} = (X^TX)^{-1} X^Ty$ and $vs^{2} = (y - X\hat{\beta})^T(y - X\hat{\beta})$.
	$\hat{\beta}$ and $s^2$ are sufficient independent statistics, such that $\hat{\beta}\sim\mathcal{N}_q(\beta,\sigma^2(X^TX)^{-1})$ and $s^2\sim \left(\frac{\sigma^2}{N-q}\right)\chi^2_{N-q}$.
	This implies
	
	\begin{equation}\label{Var1}
	\Sigma_{\hat{\beta},s^2}=
	\begin{bmatrix} \sigma^2(X^TX)^{-1} & 0\\
	0 & \frac{2\sigma^4}{N-q}  
	\end{bmatrix}
	\end{equation}
	and 
	\begin{equation}\label{alpha1}
	\alpha_{\hat{\beta},s^2}= \left[ (1/\sigma^2)(\beta-\hat{\beta})^T(X^TX) \quad (1/s^2) ((N-q)/2-1)-1/2 \right]
	\end{equation}

	The {\it{plug-in}} estimator for the optimal input would be
	
	\begin{equation}\label{plugOpt}
	\hat{\omega}^{plug}=\frac{1}{2\hat{\beta}_2}\left(\frac{w}{p}-\hat{\beta}_1\right)
	\end{equation}
	
	In addition, the application of the delta method to estimate the variance would give as a result 
	\begin{equation}\label{plugOptvar}
	\widehat{Var(\hat{\omega}^{plug})} = \frac{1}{4\hat{\beta}^{2}_{2}} \left[ \widehat{Var(\hat{\beta}_{1})} + 4 (\hat{\omega}^{plug})^2 \widehat{Var(\hat{\beta}_{2})} + 4 \hat{\omega}^{plug} \widehat{Cov(\hat{\beta}_{1}, \hat{\beta}_{2})} \right]
	\end{equation}

	On the other hand we can obtain the MELO estimate focusing directly on the inferential problem.
	We set $\epsilon = -\left(\frac{w}{p} - \beta_{1}\right) - 2\beta_{2} \hat{\omega}$ as the estimation error.
	Observe that if $\hat{\omega}$ is equal to $x^{Opt}$, the estimation error is equal to 0.\\
	
	The generalized loss function for this problem is given by
	\begin{align*}
	\mathcal{L}({\bf{g}}(\bm{\theta}),\hat{\omega}) = \epsilon^{2} & = \left[\left(\frac{w}{p} - \beta_{1}\right) - 2\beta_{2} \hat{\omega}\right]^{2} \\
	& = 4 \beta^{2}_{2} (\omega - \hat{\omega})^{2}
	\end{align*}
	
	\noindent where $\omega = {\bf{g}}(\bm{\theta}) = \frac{1}{2\beta_{2}}\left(\frac{w}{p} - \beta_{1}\right)$ and ${\bf{Q}}(\bm{\theta})=4 \beta^{2}_{2}$.\\

	Proposition \ref{proposition1} implies that the MELO estimate is
	\begin{align}\label{MELOopt}
	\hat{\omega}^{*} & = \frac{\frac{w}{p}E(\beta_2)-E(\beta_1\beta_2)}{2E(\beta_2^2)}\\ 
	& = \hat{\omega}^{plug} \left[\frac{1- \left(\frac{1}{w/p-E(\beta_1)}\right) \left(Cov(\beta_{1},\beta_{2}) / E(\beta_2)\right)}{\left(1 + Var(\beta_{2}) / E(\beta_2)^2\right)}\right]\nonumber
	\end{align}

	Using the following diffuse prior $p(\beta,\sigma) \propto 1/\sigma$, $0 < \sigma < \infty$ and $-\infty<\beta_{l}<\infty$, $l=\{1,2\}$, then the marginal posterior pdf for $\beta$ has the form of a multivariate Student-$t$ \citep{Zellner1996}:
	\begin{equation*}
	p(\beta|y, X) \propto \left\{vs^{2} + \left(\beta - \hat{\beta}\right)^T X^TX \left(\beta - \hat{\beta}\right) \right\}^{-(v+q)/2}
	\end{equation*}
	\noindent which implies that the mean of $\beta$ is $\hat{\beta}$ and its covariance matrix is $(X^TX)^{-1}vs^2/(v-2)$, $v>2$.\\
	
	We can use the previous expressions to calculate our MELO proposal (Equation {\ref{MELOopt}}), and Equations \ref{Var1} and \ref{alpha1} to obtain the frequentist variance of the MELO estimate.\\

	We set the mean deviation problem, $y_i-\bar{y}=1.5(x_i-\bar{x})-0.002(x_i^2-\overline{x^2})+u_i$, where $x_i\sim\mathcal{N}(187.5,70^2)$ and $u_i\sim\mathcal{N}(0,\sigma^2_u)$ such that $\sigma^2_u$ generates different degrees of signal to noise models $\{0.1,1,5,20\}$.
	In addition, we set the input and output prices equal to \$3,000 and \$4,000, respectively.
	This implies $x^{Opt}=187.5$.\\
	
	We perform 1,000 simulation exercises using different sample sizes (20, 50 and 500), and calculate the Mean Squared Error (MSE) and the Mean Absolute Error (MAE) for the {\it{plug-in}} approach, and the MELO using the analytical solution (Equation \ref{TheMost}), which is available in this setting, and the computational strategy of drawing from the posterior distribution (Equation \ref{TheSMost} using 10,000 iterations from a Student's $t$ distribution).\\
	
	We see from Table \ref{tab1} that the MELO outperforms the {\it{plug-in}} approach in point estimates of the optimal input; especially in the presence of noisy models and small sample sizes.
	In addition, we observe that there is no meaningful difference between the analytical and computational solutions.\\
	
	In particular, there is no clear pattern in the MSE and MAE in very noisy models as the sample size increases.
	However, we do observe that the MELO estimates outperform the {\it{plug-in}} approach in this situation.
	As the signal of the model improves, the MSE and MAE decrease as the sample size increases.
	The MSE and MAE from the MELO estimates (analytical and computational) are never worse than the {\it{plug-in}} estimates.
	However, we basically get the same outcomes using large sample sizes.
	This outcome follows from the previous asymptotic properties.
	

	\subsection{Odds ratio problem}\label{Oddsratioproblem}
	
	Setting $y_{i}$ as a dichotomous variable $\left\{0,1\right\}$ that is distributed as a Bernoulli process with parameter $p$, and assuming that the main interest is the Odds ratio, it follows that
	
	\begin{equation*}
	{\bf{g}}(\bm{\theta}) = \omega(p) = \frac{p}{1 - p}
	\end{equation*}
	\noindent where $p=P(y = 1)$.
	
	The binary probit model can be used to tackle this situation, such that  $p=P(y_{i} = 1) = \Phi (x^T_{i}\beta)$, where $\Phi(z)$ is the cumulative distribution function of the standard normal distribution evaluated at $z$.\\
	
	This model can be written with latent variables as follows:
	
	\begin{equation}\label{probit3}
	y^{*}_{i} = x^T_{i}\beta + u_{i}, \hspace{0.1cm} u_{i} \sim \mathcal{N}(0,1)
	\end{equation}
	
	\begin{equation}\label{probit4}
	y_{i} = \left\{ \begin{array}{lcc}
	0, &   if  & y^{*}_{i} \leq 0 \\
	1, &      & y^{*}_{i} > 0
	\end{array}
	\right.
	\end{equation}
	
	The likelihood function is 
	\begin{equation*}\label{likeliprobit}
	f(\beta|y,x) = \prod^{N}_{i=1} \left(\Phi (x^T_{i}\beta)^{y_{i}}  (1 - \Phi (x^T_{i}\beta))^{(1-y_{i})}\right)
	\end{equation*}
	
	Observe that in this setting there are no sufficient statistics \citep{Nelder1972}.\\ 
	
	The \textit{plug-in} estimator for the Odds ratio is
	
	\begin{equation*}
	\hat{\omega}^{plug} = \frac{\Phi (x^T_{i}\hat{\beta})}{1- \Phi (x^T_{i}\hat{\beta})}
	\end{equation*}
	
	And its variance, calculated by the delta method, is
	
	\begin{equation}\label{plugProbittvar}
	\widehat{Var(\hat{\omega}^{plug})} = \frac{\Phi(x^T_i\hat{\beta})}{N\left[1 - \Phi(x^T_i \hat{\beta})\right]^3}
	\end{equation}
	
	The loss function is given by
	\begin{align*}
	\mathcal{L}({\bf{g}}(\bm{\theta}),\hat{\omega}) = \epsilon^{2} & = \left[(1 - \Phi (x^T_{i}\beta)) \hat{\omega} - \Phi (x^T_{i}\beta)\right]^{2} 
	\end{align*}
	
	\noindent where $\omega = {\bf{g}}(\bm{\theta}) = \frac{\Phi (x^T_{i}\beta)}{1- \Phi (x^T_{i}\beta)}$ and ${\bf{Q}}(\bm{\theta}) = (1 - \Phi (x^T_{i}\beta))^{2}$.
	
	Proposition \ref{proposition1} implies that the MELO estimate is
	
	\begin{equation*}
	\hat{\omega}^{*} = \frac{E\left[(1-\Phi (x^T_{i}\beta))\Phi (x^T_{i}\beta)\right]}{E(1- \Phi (x^T_{i}\beta))^{2}}
	\end{equation*}

	Note that if $\hat{p} \rightarrow 1$, then $1-\hat{p} \rightarrow 0$, and so $\hat{\omega}^{plug} \rightarrow \infty$, while $\hat{\omega}^{*}$ can take indeterminate values of the form $0/0$.\\
	
	According to \cite{greenberg2012introduction}, using the latent variables $y^*_i$, we can write the likelihood function as
	
	\begin{align*}
	f(y_{i}|y^*_i, \beta)  = & \left[1(y_i = 0) 1(y^*_i \leq 0) + 1(y_i = 1) 1(y^*_i > 0) \right] \mathcal{N}_{N} (y^*|X\beta, I) \\
	= & \left[1(y_i = 0) 1(y^*_i \leq 0) + 1(y_i = 1) 1(y^*_i > 0)\right] \\
	\times &  \exp \left\{-\frac{1}{2} \left[vs^{2} + \left(\beta - \hat{\beta}\right)^T X^TX \left(\beta - \hat{\beta}\right)\right]  \right\}
	\end{align*}
	
	\noindent where $q=dim\{\beta\}$, $v=N-q$, $\hat{\beta} = (X^TX)^{-1} X^Ty^*$ and $vs^{2} = (y^* - X\hat{\beta})^T(y^* - X\hat{\beta})$.
	This implies that augmenting the observed binary data $y$ with the latent variable $y^*$, $\hat{\beta}$ and $vs^{2}$ are sufficient statistics, so that  \begin{equation}\label{VarProbit}
	\Sigma_{\hat{\beta},s^2}=
	\begin{bmatrix} \sigma^2(X^TX)^{-1} & 0\\
	0 & \frac{2}{N-q}  
	\end{bmatrix}
	\end{equation}
	and 
	\begin{equation}\label{alphaProbit}
	\alpha_{\hat{\beta},s^2}= \left[ (\beta-\hat{\beta})^T(X^TX) \quad (1/s^2) ((N-q)/2-1)-1/2 \right]
	\end{equation}

	Assuming a normally distributed prior for $\beta$, the posterior distributions of $\beta$ and $y^*$ 
	are
	
	\begin{equation}\label{posteriorprobit}
	\pi(\beta, y^*|y) \propto \prod^{N}_{i=1} \left\{1(y_i = 0) 1(y^*_i \leq 0) + 1(y_i = 1) 1(y^*_i > 0)\right\} \mathcal{N}_N (y^*|X\beta, I) \mathcal{N}_q (\beta|\beta_0, B_0)      
	\end{equation}
	
	Therefore,
	
	\begin{algorithm}[!h]
		\caption{Bayesian Probit Model}
		\label{alg:BP}
		\begin{algorithmic}[1]
			\State Choose a starting value $\beta^{(0)}$
			\State At the $g$th iteration, draw
			\begin{equation*}
			y_i^* \sim \left\lbrace
			\begin{array}{ll}
			\mathcal{T}\mathcal{N}_{(-\infty,0)}(x_i^T\beta^{(g-1)},1), &  y_i=0\\
			\mathcal{T}\mathcal{N}_{(0,\infty)}(x_i^T\beta^{(g-1)},1), & y_i=1
			\end{array}
			\right.
			\end{equation*}
			\State $\beta^{(g)}\sim\mathcal{N}_q(\hat{\beta}^{(g)},B_1)$, where $B_1=(X^TX+B_0^{-1})$ and $\hat{\beta}^{(g)}=B_1(X^Ty^{*(g)}+B_0^{-1}\beta_0)$.
			
		\end{algorithmic}
	\end{algorithm}

	Consider the following setting.
	
	\begin{equation}
	y^{*}_{i} = 0.5 + 0.8 x_{1,i}  -  1.2 x_{2,i} + \mu_{i} 
	\end{equation}
	
	We simulate the data set $x_{1}$, $x_{2}$ and the stochastic errors from standard normal distributions, and perform 1,000 simulation exercises using four different sample sizes: 20, 50, 500 and 1,000.\\

	Tables \ref{Table2} and \ref{Table3} show the mean errors of our simulation exercises.
	In particular, we perform two different evaluations for the Odds ratio, $x=(1,1,1)$ and $x=(1,0,0)$ using Algorithm \ref{alg:BP} while setting $B_0=10,000 \ diag\left\{1,1,1\right\}$ and $\beta_0=\left[0,0,0\right]$ with 25,000 iterations and a burn-in equal to 5,000.
	We see from these tables that the range of variability of the different measures of the MELO approach is lower than for the {\it{plug-in}} approach.
	We observe that when the sample size is small, the differences are remarkable, especially when $x=(1,1,1)$, that is, when the data is less informative (noisy) due to regressors not being located in their population means ($x=(1,0,0)$).
	We obtain similar results for both approaches as the sample sizes increases.

	\subsection{Portfolio selection}
	
	One strategy for active portfolio management looks for finding the asset weights that maximize the Sharpe ratio, that is, the mean portfolio return per unit of risk.

	\begin{equation*}
	\max_{{\bf{w}}\in \mathcal{R}^L} \frac{{\bf{w}}^T{{\tilde{\mu}}}}{({\bf{w}}^T\tilde{\Sigma} {\bf{w}})^{1/2}}
	\hspace{0.5cm}
	\mbox{s.t}
	\hspace{0.5cm}
	{\bf{w}}^T{\bf{1}}=1
	\end{equation*}
	
	\noindent where $\tilde{\mu}$ is the mean vector of the asset's excess returns in the investment period (say $\tau$), $\tilde{\Sigma}$ is its covariance matrix, and ${\bf{1}}$ is a vector of ones.\\
	
	The solution of the previous problem gives the well known tangent portfolio, that is,
	
	\begin{equation}\label{optweight}
	{\bf{w}}^{Opt}=\frac{\tilde{\Sigma}^{-1}{\tilde{\mu}}}{{\bf{1}}^T\tilde{\Sigma}^{-1}\tilde{\mu}}
	\end{equation}
	
	As we can see from Equation \ref{optweight}, the final aim of the inferential problem is a rational function of the parameters of the asset's excess returns.\footnote{Given $A_{d\times d}$ invertible, then there exists a polynomial $p$, such that $A^{-1}=p(A)$.}
	
	The standard financial literature assumes that the asset's excess returns are jointly normally distributed, i.e., $r_t\sim\mathcal{N}_d(\mu, \Sigma)$ for $t=1,2,\ldots,T$, where the excess returns are serially independent.\\
	
	Now put ${\bf{R}}=({\bf{r_1}}, {\bf{r_2}}, \ldots, {\bf{r_L}})$ a $T\times L$ matrix of observations on $L$ asset excess returns.
	Then we can write the following model for the excess returns:
	
	$${\bf{R}}={\bf{1}}\mu^T+e$$
	
	\noindent where $e=(e_1, e_2, \ldots, e_L)$ is an $T\times L$ matrix of unobserved random disturbances.
	The rows of $e$ are independently distributed, which precludes any auto or serial correlation of disturbance terms, each with an $L$-dimensional normal distribution with zero mean vector and positive definite $L\times L$ covariance matrix $\Sigma$.\\
	
	The likelihood of this model is
	
	$$f(\mu,\Sigma|{\bf{R}})\propto |\Sigma|^{-T/2} exp\left\{-\frac{1}{2}tr(S\Sigma^{-1})-\frac{1}{2T}tr((\mu-\hat{\mu})(\mu-\hat{\mu})^T\Sigma^{-1})\right\}$$
	
	\noindent where $\hat{\mu}$ is the sample mean vector and $S=({\bf{R}}-{\bf{1}}\hat{\mu}^T)^T({\bf{R}}-{\bf{1}}\hat{\mu}^T)$.
	$\hat{\mu}$ and $S$ are sufficient statistics, such that $\hat{\mu}\sim\mathcal{N}_L(\mu,\Sigma)$ and $S\sim \mathcal{W}_L(T-1,\Sigma)$.
	$\hat{\mu}$ and $S/(T-1)$ are consistent estimators for $\mu$ and $\Sigma$.
	Then,
	
	\begin{equation}
	\alpha_{\hat{\mu},\hat{\Sigma}}= \left[ (\mu-\hat{\mu})^T\Sigma^{-1} \quad vec\left(\left(\frac{T-1-L-1}{2}\right)S^{-1}-\frac{1}{2}\Sigma^{-1}\right)^T \right]
	\end{equation}
	
	\begin{equation}
	\Sigma_{\hat{\bm{\theta}}}=
	\begin{bmatrix} \Sigma & 0\\
	0 & \Sigma_S  
	\end{bmatrix}
	\end{equation}
	\noindent where $Var(S_{ij})=(T-1)(\sigma_{ij}^2+\sigma_{ii}\sigma_{jj})$ and $Cov(S_{ij},S_{kl})=(T-1)(\sigma_{ik}\sigma_{jl}+\sigma_{il}\sigma_{jk})$.\\
	
	The {\it{plug-in}} estimator for the tangent portfolio is
	
	\begin{equation}\label{optweightplug}
	\hat{{\bf{w}}}^{plug}=\frac{\hat{\Sigma}^{-1}{\hat{\mu}}}{{\bf{1}}^T\hat{\Sigma}^{-1}\hat{\mu}}
	\end{equation}
	
	On the other hand we can obtain the MELO estimate focusing directly on the inferential problem.
	We set $\epsilon=({\bf{1}}^T\tilde{\Sigma}^{-1}{\tilde{\mu}}) {\hat{\omega}}-\tilde{\Sigma}^{-1}{{\tilde{\mu}}}$ as the estimation error.
	Observe that if $\hat{\omega}$ is equal to ${\bf{w}}^{Opt}$, the estimation error is equal to 0.\\
	
	Given ${\bf{g}}(\bm{\theta})=\frac{\tilde{\Sigma}^{-1}{\tilde{\mu}}}{{\bf{1}}^T\tilde{\Sigma}^{-1}{\tilde{\mu}}}$, the generalized loss function for this problem is given by $\mathcal{L}({\bf{g}}(\bm{\theta}),\hat{\omega})=\epsilon^T\epsilon$ and $E(\mathcal{L})=E_{\pi(\mu,\Sigma|{\bf{R}})}\epsilon^T\epsilon$, ${\bf{Q}}(\bm{\theta})=({\bf{1}}^T\tilde{\Sigma}^{-1}{\mu})^2$.
	Despite the fact that the expected value should be based on information up to the investment period ($T+\tau$), we only have information up to $T$, so the expected value is conditioned on $\textbf{R}$.
	However, informative priors can be based on experts' views of the investment period.\\
	
	Proposition \ref{proposition1} implies that the MELO estimate is
	\begin{align}\label{MELOoptWeight}
	\hat{\omega}^{*} & =\frac{E_{\pi(\mu,\Sigma|{\bf{R}})}(({\bf{1}}^T\tilde{\Sigma}^{-1}{\tilde{\mu}})\tilde{\Sigma}^{-1}{\tilde{\mu}})}{E_{\pi(\mu,\Sigma|{\bf{R}})}({\bf{1}}^T\tilde{\Sigma}^{-1}{\tilde{\mu}})^2}   
	\end{align}

	Using the diffuse prior $\pi(\mu,\Sigma)=\pi(\mu)\pi(\Sigma)\propto |\Sigma|^{-(L+1)/2}$, the conditional posterior distribution for the mean vector of asset excess returns is $\mu|\Sigma,{\bf{R}}\sim\mathcal{N}_L(\hat{\mu},\Sigma/T)$, and the marginal distribution for the covariance matrix is $\Sigma|{\bf{R}}\sim\mathcal{I}\mathcal{W}_L(T-1,S)$ \citep{Zellner1996}.
	Therefore, we can use a Gibbs sampling algorithm to obtain a computational solution for our MELO estimate.\footnote{Observe that following the {\textit{concentional}} Bayesian portfolio selection, which is based on the predictive distribution of the excess returns in the investment period, we have $\tilde{\mu}=\hat{\mu}$ and $\tilde{\Sigma}=\frac{\left(\tau+\frac{1}{T}\right)(T-1)}{T+\tau-2-L}\hat{\Sigma}$.
		The term $\frac{\left(\tau+\frac{1}{T}\right)(T-1)}{T+\tau-2-L}$ cancels out in Equation \ref{MELOoptWeight}.}
	
	We set $\mu_l\sim\mathcal{U}(-0.2,0.2)$, $l=1,2,\dots,L$, and generate $\Sigma$ such that it is semidefinite positive.
	We set four different scenarios of portfolio selection: $L=\left\{10, 25, 50, 100\right\}$ assets, and two sample sizes: $T=\left\{120, 240\right\}$ periods.
	We perform 100 simulations for each of the 8 settings, so that ${\bf{R}}\sim\mathcal{N}(\mu,\Sigma)$.\\
	
	We estimate the sample mean and covariance matrix to calculate the optimal weights using the {\it{plug-in}} approach (Equation \ref{optweightplug}), and the Gibbs sampling algorithm with 1,000 iterations to calculate our MELO proposal (Equation \ref{MELOoptWeight}).
	Then, we obtain the MSE and MAE using the population parameters (Equation \ref{optweight}), and the two estimators.
	We can see in Tables \ref{MSE} and \ref{MAE} the outcomes of our simulation exercises.
	In particular, the mean of the MSE and MAE associated with the MELO is always lower than the {\it{plug-in}} approach; there are remarkable improvements of our proposal when the number of assets in the portfolio selection problem is small.
	In the latter cases, the range of variability in the MSE and MAE using the {\it{plug-in}} approach is enormous compared with the MELO approach.
	
	
	\subsection{Structural supply--demand model}\label{structural}
	Assume the following structural supply--demand model:
	\begin{align}\label{estructural}
	q^d_i&=\beta_0+\beta_1p_i+\beta_2z_{1i}+\mu_{di}\\
	q^s_i&=\alpha_0+\alpha_1p_i+\alpha_2z_{2i}+\mu_{si}
	\end{align}
	\noindent where $q^d_i$ and $q^s_i$ are demand and supply functions, $p_i$ is the price, $z_{1i}$ and $z_{2i}$ exogenous regressors, and $\mu_{di}$ and $\mu_{si}$ stochastic errors, $i=1,2,\dots,N$.\\
	
	The equilibrium condition equates demand and supply, that is, $q^s=q^d$.
	the structural parameters are the main concern of the econometric inferential problem.
	
	Equation \ref{estructural} cannot be directly estimated due to endogeneity issues.
	So, it is necessary to obtain the reduced form system
	\begin{align*}
	q_i&=\pi_0+\pi_1z_{1i}+\pi_2z_{2i}+e_{qi}\\
	p_i&=\gamma_0+\gamma_1z_{1i}+\gamma_2z_{2i}+e_{pi}
	\end{align*}
	
	\noindent which can be written as ${\bf{Y}}={\bf{X}}B+{\bf{U}}$, where ${\bf{Y}}=\left[q \quad p\right]$, an $N\times 2$ matrix of observations on quantities and prices, ${\bf{X}}$ is an $N\times 3$ matrix of a vector of ones, and the two independent variables ($z_1$ and $z_2$), with rank $3$, $B=\left[\pi\quad\gamma\right]$ is a $3\times 2$ matrix of regressions parameters from the reduced form, and ${\bf{U}}=\left[e_q\quad e_p\right]$ is a $N\times 2$ matrix of unobserved stochastic errors.
	We assume that the rows of $U$ are independently distributed, each with a 2-dimensional normal distribution with zero mean vector and positive definite $2\times 2$ covariance matrix $\Sigma$.\\
	
	The likelihood function of this system is 
	$$f(B,\Sigma|{\bf{Y}},{\bf{X}})\propto |\Sigma|^{-T/2} exp\left\{-\frac{1}{2}tr({\bf{S}}\Sigma^{-1})-\frac{1}{2}tr((B-\hat{B})^T({\bf{X}}^T{\bf{X}})(B-\hat{B})\Sigma^{-1})\right\}$$
	
	\noindent where $\hat{B}=({\bf{X}}^T{\bf{X}})^{-1}{\bf{X}}^T{\bf{Y}}$ a matrix of least squares quantities, and ${\bf{S}}=({\bf{Y}}-{\bf{X}}\hat{B})^T({\bf{Y}}-{\bf{X}}\hat{B})$.
	$\hat{B}$ and $S$ are sufficient statistics, such that $vec(\hat{B})=\left[\hat{\pi}^T \quad \hat{\gamma}^T\right]^T\sim\mathcal{N}_6(vec(B),\Sigma\otimes ({\bf{X}}^T{\bf{X}})^{-1})$ and ${\bf{S}}\sim \mathcal{W}_2(N-3,\Sigma)$.
	$\hat{B}$ and ${\bf{S}}/(N-3)$ are consistent estimators for $B$ and $\Sigma$.\\
	
	Then,
	\begin{equation*}
	\alpha_{\hat{\beta},\hat{\Sigma}}= \left[ (\beta-\hat{\beta})^T(\Sigma^{-1}\otimes({\bf{X}}^T{\bf{X}})) \quad vec\left(\left(\frac{N-3-2-1}{2}\right){\bf{S}}^{-1}-\frac{1}{2}\Sigma^{-1}\right)^T \right]
	\end{equation*}
	\noindent and
	\begin{equation*}
	\Sigma_{\hat{\beta},\hat{\Sigma}}=
	\begin{bmatrix} \Sigma\otimes ({\bf{X}}^T{\bf{X}})^{-1} & 0\\
	0 & \Sigma_S  
	\end{bmatrix}
	\end{equation*}
	\noindent where $\beta=vec(B)$, $\hat{\beta}=vec(\hat{B})$,  $Var(S_{ij})=(N-3)(\sigma_{ij}^2+\sigma_{ii}\sigma_{jj})$ and $Cov(S_{ij},S_{kl})=(N-3)(\sigma_{ik}\sigma_{jl}+\sigma_{il}\sigma_{jk})$.\\

	The relation between the structural parameters, which are the main concern of the econometric inferential problem, and the reduced form parameters is given by the following system of equations:
	
	\begin{equation}\label{structu}
	\begin{bmatrix}
	\beta_1\\
	\beta_2\\
	\alpha_1\\
	\alpha_2\\
	\end{bmatrix} = \begin{bmatrix}
	\pi_2/\gamma_2\\
	\pi_1-\gamma_1\pi_2/\gamma_2\\
	\pi_1/\gamma_1\\
	\pi_2-\gamma_2\pi_1/\gamma_1\\
	\end{bmatrix}
	\end{equation}
	
	There are different alternatives for obtaining the structural parameters from the reduced form.
	In this setting, which is an exactly identified model, the point estimates using the ILS ({\it{plug-in}} approach), 2SLS, or 3SLS, give the same results.\\
	
	We set the vector of errors to be
	\begin{equation*}
	\begin{bmatrix}
	\epsilon_1\\
	\epsilon_2\\
	\epsilon_3\\
	\epsilon_4\\
	\end{bmatrix} = \begin{bmatrix}
	\gamma_2(\hat{\omega}_1-\beta_1)\\
	\gamma_2(\hat{\omega}_2-\beta_2)\\
	\gamma_1(\hat{\omega}_3-\alpha_1)\\
	\gamma_1(\hat{\omega}_4-\alpha_2)\\
	\end{bmatrix} 
	\end{equation*}
	
	The loss function is $\mathcal{L}({\bf{g}}(\bm{\theta}),\hat{\omega}) = \epsilon^{T}\epsilon$, which implies ${\bf{Q}}=diag(\gamma_2^2,\gamma_2^2,\gamma_1^2,\gamma_1^2)$.
	As a consequence, the MELO is given by the following set of simultaneous equations.
	
	\begin{equation*}
	\begin{bmatrix}
	\hat{\omega}_1^{*}\\
	\hat{\omega}_2^{*}\\
	\hat{\omega}_3^{*}\\
	\hat{\omega}_4^{*}\\
	\end{bmatrix} = \begin{bmatrix}
	E(\pi_2\gamma_2)/E(\gamma_2^2)\\
	(E(\pi_1\gamma_2^2)-E(\gamma_1\gamma_2\pi_2))/E(\gamma_2^2)\\
	E(\pi_1\gamma_1)/E(\gamma_1^2)\\
	(E(\pi_2\gamma_1^2)-E(\gamma_1\gamma_2\pi_1))/E(\gamma_1^2)\\
	\end{bmatrix}
	\end{equation*}
	
	Observe that the different components of the MELO estimates are independent.
	This is due to the structure of the weighting matrix: it is a diagonal matrix.
	So, we can focus our effort on the specific structural parameters of interest.\\
	
	Using the diffuse prior $\pi(B,\Sigma)=\pi(B)\pi(\Sigma)\propto |\Sigma|^{-(2+1)/2}$, the conditional posterior distribution for the mean vector is $vec(B)|\Sigma,{\bf{Y}},{\bf{X}}\sim\mathcal{N}_6(vec(\hat{B}),\Sigma\otimes ({\bf{X}}^T{\bf{X}})^{-1})$, and the marginal distribution for the covariance matrix is $\Sigma|{\bf{Y}},{\bf{X}}\sim\mathcal{I}\mathcal{W}_2(N-3,S)$ \citep{Zellner1996}.
	Therefore, we can use a Gibbs sampling algorithm to obtain a computational solution of our MELO estimate.\\ 
	We consider the following structural supply--demand model:
	\begin{align}\label{estructuralsim}
	q^d_i&= 0.2-0.8p_i+1.5z_{1i}+\mu_{di}\\
	q^s_i&= -0.5+1.2p_i-z_{2i}+\mu_{si}
	\end{align}
	\noindent which implies the following reduced form model:
	\begin{align}\label{reducedsim}
	q_i&= 0.35+0.75z_{1i}+0.50z_{2i}+e_{qi}\\
	p_i&= -0.08+0.9z_{1i}-0.4z_{2i}+e_{pi}
	\end{align}
	
	We simulate $z_{1i}$ and $z_{2i}$ from standard normal distributions, and the stochastic errors from the reduced system as independent variables with mean zero and standard deviation such that the signal to the noise ratio in the reduced equations are simultaneously equal to 0.1, 0.5, 1 and 5.
	We know that from a theoretical point of view this is a mistake since there is a correlation between the stochastic errors in the reduced form system.
	However, we follow this setting to have independence between the equations, and as a consequence we know that the marginal posterior distributions of each equation are independent multivariate Student's $t$ distributions.
	This implies that
	
	\begin{equation*}
	\begin{bmatrix}
	\hat{\omega}_1^{*}\\
	\hat{\omega}_2^{*}\\
	\hat{\omega}_3^{*}\\
	\hat{\omega}_4^{*}\\
	\end{bmatrix} = \begin{bmatrix}
	E(\pi_2)E(\gamma_2)/(Var(\gamma_2)+(E(\gamma_2))^2)\\
	E(\pi_1)-(E(\pi_2)(Cov(\gamma_1,\gamma_2)+E(\gamma_1)E(\gamma_2))/(Var(\gamma_2)+(E(\gamma_2))^2)\\
	E(\pi_1)E(\gamma_1)/(Var(\gamma_1)+(E(\gamma_1))^2)\\
	E(\pi_2)-(E(\pi_1)(Cov(\gamma_1,\gamma_2)+E(\gamma_1)E(\gamma_2))/(Var(\gamma_1)+(E(\gamma_1))^2)\\
	\end{bmatrix}
	\end{equation*}
	
	\noindent and so we can compare the analytical and computational versions of the MELO estimates with the frequentist competing alternative.
	In particular, we use 50,000 iterations for the Gibbs sampling algorithm used to calculate the computational MELO.
	\\
	
	We can see in Table \ref{Table4} the mean errors associated with 1,000 simulations exercises using five different sample sizes: 20, 50, 100, 1,000 and 20,000.
	We observe from this table the same pattern as in the previous simulations exercises.
	The MELO estimates outperform 2SLS in terms of point estimates,
	especially in situations characterized by noisy models and small sample sizes.
	However, we always get the same performance with a sample size equal to 20,000 or a signal to noise ratio equal to 5.
	The performance of the three approaches improves as the sample size increases as well as when the signal to noise increases.
	In general, the MELO estimates are never worse than those of the 2SLS, and the analytical and computational solutions have the same performance.
	
	
	\section{Applications}\label{App}
	
	\subsection{Experimental broiler input--output}
	This is the broiler input--output example presented by \cite{Judge1988}.
	In particular, the average weight of an experimental lot of broilers and their corresponding levels of average feed consumption was tabulated over the time period in which they changed from baby chickens to mature broilers ready for market.\\
	
	Given the setting of the optimal input problem in subsection \ref{subsec1}, the dataset in Table 5.3 from \cite{Judge1988}, and taking into account that broilers are 30 cents per pound and feed is 6 cents per pound, the optimal level of feed input is 13.74 with a standard deviation equal to 1.89 using the {\it{plug-in}} approach (Equations \ref{plugOpt} and \ref{plugOptvar}), whereas the optimal input point estimate using the MELO approach, both analytical (Equation \ref{MELOopt}) and computational (Equation \ref{TheSMost} using 10,000 iterations), is 13.14. The standard deviations is 1.46, that is, reductions of 22\%.
	These figures are calculated using Equation \ref{alpha1} with Corollary \ref{corollary}, and Equation \ref{Var1}, with Proposition \ref{proposition5} in the case of the analytical and computational approaches.
	Despite the fact that the coefficient of determination in this example is very high ($R^2=0.98$), we observe differences between the optimal weight estimates.
	In addition, the frequentist variability of the optimal weight using the MELO estimates are lower than the one using the {\it{plug-in}} approach.
	
	\subsection{Space Shuttle Challenger}
	
	In 1986, the space shuttle Challenger exploded during take off, killing the seven astronauts aboard.
	The explosion was the result of an O-ring failure, a splitting of a ring of rubber that seals the parts of the ship together, due to the unusually cold weather ($31^{o}F$, i.e., $0^{o}C$) at the time of launch \citep{dalal1989risk}.\\
	
	We calculated the Odds ratio at $45^{o}F$ and $69.56^{o}F$ (mean sample temperature) for a sample of 23 observations provided by \cite{robert2004monte} taking into account the theoretical structure of subsection \ref{Oddsratioproblem}.
	Using the {\it{plug-in}} approach, the probability of failure is 0.996 at $45^{o}F$, therefore the odds ratio estimate is 283.644 with a standard deviation of 999.596.
	The Odds ratio using MELO is 2.585 in the case of the computational approach (Algorithm \ref{alg:BP} setting $B_0=10,000 \ diag\left\{1,1,1\right\}$ and $\beta_0=\left[0,0,0\right]$ with 25,000 iterations and a burn-in equal to 5,000).
	Observe that the implicit probabilities of the Odds ratio in the MELO approach is 0.721.
	However, if the main objective of the statistical inference is the probability, that is, ${\bf{g}}(\bm{\theta})=\Phi(x^T\beta)$, which implies ${\bf{Q}}(\bm{\theta})=1$, and $\hat{\omega}^{*}=E(\Phi (x^T_{i}\beta))$, we have point estimates equal to 0.964 using the computational MELO. This highlights a remarkable characteristic of our approach; the estimate depends drastically on the main objective of the inferential situation.\\
	
	Regarding the frequentist variability of the MELO, we get 2.917 using the computational approach.
	Observe that in this case, the components associated with Corollary \ref{corollary} depend on the iteration $g$, so we calculate the mean values over all these components to obtain this figure. Observe that there is a huge difference using the delta method (999.596).\\  
	
	The failure probability is 0.266 at $69.56^{o}F$ using the {\it{plug-in}} approach.
	This implies an Odds ratio equal to 0.363 with standard deviation equal to 0.171.
	The Odds ratio using the computational MELO is 0.345 with a standard deviation equal to 0.258.
	We get similar point estimates using the central point in the distribution of regressors.
	In this case, the standard deviation of the {\it{plug-in}} is lower than for the MELO approaches (33\%).\\
	
	The message here is that in the case of evaluating a point in the extreme of the distribution of the regressors, that is, when the sample information is not precise (noisy), it is much better to use the MELO approach.
	On the other hand, it makes sense to use the {\it{plug-in}} approach.

	\subsection{Colonial origins of development}
	\cite{Acemoglu2001} analyze the effect of property rights on economic growth.
	They exploit the variability in European settlers' mortality rates during the time of colonization to find the causal effect of protection against expropriation on economic performance.
	They use 2SLS to accomplish this task.
	We can write their setting in the following structural system,
	\begin{align*}
	Log(pcGDP)_i =& \beta_{0}+\beta_1PAER_i+\mu_{1i}\\
	PAER_i=&\alpha_{0}+\alpha_1Log(pcGDP)_i+\alpha_{2}log(Mort)_i+\mu_{2i}	
	\end{align*}
	\noindent where $pcGDP$, $PAER$ and $Mort$ are the per capita GDP in 1995, the average index of protection against expropriation between 1985 and 1995 (Political Risk Services), and settler mortality rate during the time of colonization (see \cite{Acemoglu2001} for details), respectively.	The reduced form model is
	\begin{align*}
	Log(pcGDP)_i =& \pi_{0}+\pi_1log(Mort)_i+e_{1i}\\
	PAER_i =& \gamma_{0}+\gamma_1log(Mort)_i+e_{2i}	
	\end{align*}
	
	The first structural equation is exactly identified provided that $\alpha_{2}\neq 0$, whereas the second structural equation is sub-identified.\\
	
	We define the estimation error as $\epsilon=\gamma_1(\hat{\omega}-\beta_1)$, where $\beta_1=\pi_1/\gamma_1$, then ${\bf{Q}}(\bm{\theta})=\gamma_1^2$.\\
	
	We find the MELO estimates, and their frequentist variability, using the same ideas of subsection \ref{structural}.
	The outcomes can be seen in Table \ref{origins}, where we reproduce the outcomes from \cite{Acemoglu2001}, Table IV (page 1386), columns 1, 3, 5 and 9.\\
	
	We can see from Table \ref{origins} that the standard errors of our approach are always less than the standard errors from 2SLS.
	We obtain more efficiency gains in noisier models, for instance column (3), where the coefficient of determination is the lowest ($R^2=0.13$).
	In general, the MELO estimates of the effects of property rights on economic performance are lower than the 2SLS estimates. 
	
	\subsection{Openness and inflation}
	\cite{Romer1993} analyzes the effect of openness on inflation.
	In particular, he shows there is a strong and robust negative link between inflation and openness using 2SLS, where he uses the logarithm of the country's land area as an instrument of openness.
	His model can be written as a structural system of equations:
	\begin{align*}
	Inf_i =& \beta_{0}+\beta_1Open_i+\beta_{2}log(pinc_i)+\beta_{3}D_i+\mu_{1i}\\
	Open_i=&\alpha_{0}+\alpha_1Inf_i+\alpha_{2}log(pinc_i)+\alpha_{3}log(land_i)+\alpha_{4}D_i+\mu_{2i}		
	\end{align*}
	\noindent where $Inf_i$, $Open_i$, $pinc_i$,  $D_i$ and $land_i$ are the inflation rate, openness, which is measured as the ratio of imports to GDP,  real per capita income, and data dummies for the alternative measures of openness and inflation, and land area, respectively.\\
	
	The first structural equation is exactly identified provided that $\alpha_{3}\neq 0$, whereas the second structural equation is sub-identified.
	The reduced form of this model is
	\begin{align*}
	Inf_i =& \pi_{0}+\pi_1log(pinc_i)+\pi_2log(land_i)+\pi_3D_i+e_{1i}\\
	Open_i =& \gamma_{0}+\gamma_1log(pinc_i)+\gamma_2log(land_i)+\gamma_3D_i+e_{2i}	
	\end{align*}
	
	We define the estimation error as follows:
	\begin{equation*}
	\begin{bmatrix} \epsilon_1 \\
	\epsilon_2  
	\end{bmatrix} =
	\begin{bmatrix} \gamma_2(\hat{\omega}_1-\beta_1) \\
	\gamma_2(\hat{\omega}_2-\beta_2)  
	\end{bmatrix}
	\end{equation*}
	\noindent where $\beta_1=\pi_2/\gamma_2$, $\beta_2=\pi_1-\frac{\pi_2}{\gamma_2}\gamma_1$, then ${\bf{Q}}(\bm{\theta})=diag(\gamma_2^2,\gamma_2^2)$.\\
	
	We can find the MELO estimates, and their frequentist variability, using the same ideas of subsection \ref{structural}.
	In particular, we have that the structural or causal effect of openness on inflation is -1.252 with a standard error equal to 0.407, and the effect of per capita income on inflation is equal to -0.045 with a standard error of 0.061 (using the computational approach with 10,000 iterations).
	The analogous estimates using 2SLS are -1.260 (0.414) and -0.045 (0.061) for openness and income, respectively.\\
	
	Despite the fact that MELO and 2SLS are based on completely different frameworks, we practically do not get any differences between these estimates in this application.
	The reason is that the coefficient of determination in the first stage is equal to 0.48, and there are 100 degrees of freedom.
	This implies that the signal to noise ratio in the first stage is approximately 1, and given 100 d.f., we are basically replicating the outcomes of our simulation exercise in subsection \ref{structural} (see Table \ref{Table4}, Signal/Noise=0.5 and Sample size=100).\\	
	\section{Concluding remarks}
	Many times the main concern of an econometric inference is associated with rational functions of parameters.
	Our approach tackles directly this issue based on a Bayesian decision theory framework, which allows thinking about the whole inferential situation.
	Our proposal seems to improve the econometric inference in situations characterized by small sample sizes or noisy models.
	So, our MELO proposal can be used in situations where getting observations can be a  difficult task due to data limitations, for instance, expensive experimental designs or availability restrictions, and/or situations where the models are very noisy, for instance, very weak instruments.
	But, if there is a moderate sample size and/or the models are very informative, it is better to use the commonly used alternatives, due to the availability of the appropriate software for them.\\
	
	However, we must acknowledge that our approach is based on rational functions and sufficient statistics. Future research should explore relaxing these assumptions.
	\clearpage
	\bibliography{BiblioV2}   
	\bibliographystyle{apalike} 
\clearpage		
	\begin{table}[H]
		\centering
		\caption{Tangency portfolio: Mean Squared Error}\label{MSE}
		\scalebox{0.87}{
			{\footnotesize{
					\begin{tabular}{ccccccccc}
						\hline\hline
						\multicolumn{9}{c}{Mean Squared Error }\\\hline \hline
						\multicolumn{9}{c}{$ Assets=10$ }\\ \hline
						Method&	Sample size	&Min	&1st Qu.	& Median &	Mean	&3rd Qu.&	Max&	Range\\ \hline
						Plug-in &120 &0.0307	&0.0951&	0.2266&	12.5700&	0.9281&	627.6000&	627.5693
						\\ 
						MELO 	&120 &0.0126&	0.0627&	0.0947&	0.1089&	0.1261&	0.2787&	0.2661
						\\
						Plug-in &240 &0.0200&	0.0680&	0.1970&	41.8200&	1.2060&	3,306.0000&	3,305.9800
						\\
						MELO 	&240 &0.0185&	0.0606&	0.0767&	0.0949&	0.1138&	0.4032&	0.3847
						\\
						\hline\hline
						\multicolumn{9}{c}{$ Assets=25$ }\\ \hline
						Plug-in &120 &0.0048&	0.0116&	0.0221&	0.7194&	0.1083&	36.2200&	36.2152
						\\ 
						MELO 	&120 &0.0026&	0.0067&	0.0097&	0.0112&	0.0141&	0.0454&	0.0428
						\\
						Plug-in &240 &0.0027&	0.0057&	0.0102&	0.0573&	0.0205&	1.3780&	1.3753
						\\
						MELO 	&240 &0.0024&	0.0049&	0.0072&	0.0083&	0.0106&	0.0345&	0.0321
						\\
						\hline\hline
						\multicolumn{9}{c}{$ Assets=50$ }\\ \hline
						Plug-in &120 &0.0150&	0.0298&	0.0627&	20.1600&	0.2210&	1,618.0000&	1,617.9850
						\\ 
						MELO 	&120 &0.0133&	0.0266&	0.0320&	0.0421&	0.0480&	0.1242&	0.1109
						\\
						Plug-in &240 &0.0062&	0.0191&	0.0282&	5.7330&	0.1950&	131.1000&	131.0938
						\\
						MELO 	&240 &0.0079&	0.0184&	0.0213&	0.0340&	0.0377&	0.1024&	0.0945
						\\
						\hline\hline
						\multicolumn{9}{c}{$ Assets=100$ }\\ \hline
						Plug-in &120 &1.5100E-05&	4.7920E-05&	7.4650E-05&	1.2590E-04&	1.2210E-04&	1.8410E-03&	0.0018
						\\ 
						MELO 	&120 & 1.1900E-05&	4.7780E-05&	7.4800E-05&	1.2380E-04&	1.2110E-04&	1.7300E-03&	0.0017
						\\
						Plug-in &240 &2.2100E-06&	5.8250E-06&	8.8950E-06&	1.3330E-05&	1.6180E-05&	6.3400E-05&	0.0001
						\\
						MELO 	&240 &2.1900E-06&	5.9950E-06&	9.0150E-06&	1.3290E-05&	1.5970E-05&	6.2100E-05&	0.0001
						\\ \hline\hline
				\end{tabular}	}
		}}	
	\end{table}
	
	\begin{table}[H]
		\centering
		\caption{Optimal input: Mean Errors}\label{tab1}
		\scalebox{0.67}{
			\begin{tabular}{clccc}
				\hline\hline
				Signal/Noise &	Method&	Sample size&	MSE&	MAE \\
				\hline
				\multirow{9}{*}{0.1} & Plug-in & 20 &1,938,902.07&	188.63 \\
				&Analytical MELO &20  & 1,155.06&	13.25 \\
				&Computational MELO &20  & 1,153.36&	13.24 \\
				&Plug-in &50  & 1,376,539.53&	197.18\\
				&Analytical MELO &50  &5,610.28	&15.70 \\
				&Computational MELO &50  & 5,571.76&	15.67 \\
				&Plug-in & 500 & 30,593,621.65	&337.20 \\
				&Analytical MELO & 500 &3,808.44&	16.84 \\
				&Computational MELO & 500 & 3,807.83	&16.85 \\ \hline
				\multirow{9}{*}{1}& Plug-in & 20 &426,199.12&	146.97 \\
				&Analytical MELO &20  &323.90&	12.74 \\
				&Computational MELO &20  &323.80&	12.75\\
				&Plug-in &50  &13,846.97	&40.31 \\
				&Analytical MELO &50  & 346.66	&13.83 \\
				&Computational MELO &50  &346.26	&13.82 \\
				&Plug-in & 500 &124.33	&7.46 \\
				&Analytical MELO & 500 &116.58	&7.17 \\
				&Computational MELO & 500 &116.64&	7.17 \\ \hline
				\multirow{9}{*}{5} & Plug-in & 20 &189.93&	9.33 \\
				&Analytical MELO &20  &112.92&	8.21\\
				&Computational MELO &20  &112.88&	8.21 \\
				&Plug-in &50  &26.75	&4.03\\
				&Analytical MELO &50  &26.05&	3.97 \\
				&Computational MELO &50  &26.05&	3.97 \\
				&Plug-in & 500 & 4.55&	1.44  \\
				&Analytical MELO & 500 &4.54	&1.43 \\
				&Computational MELO & 500 & 4.54	&1.43 \\ \hline
				\multirow{9}{*}{20} & Plug-in & 20 &7.06& 2.10  \\
				&Analytical MELO &20& 7.00&	2.09\\
				&Computational MELO &20 &7.00&	2.09  \\
				&Plug-in &50  & 1.61& 0.99     \\
				&Analytical MELO &50  &1.61& 0.99  \\
				&Computational MELO &50  &1.61& 0.99  \\
				&Plug-in & 500 & 0.28 &0.36  \\
				&Analytical MELO & 500 & 0.28 &0.36 \\
				&Computational MELO & 500 & 0.28 &0.36 \\ 
				\hline\hline
			\end{tabular} 
		}
	\end{table}	
	
	\input{Table2}

	\input{Table3}

	\begin{table}[H]
		\centering
		\caption{Tangency portfolio: Mean Absolute Error}\label{MAE}
		{\small{
				\begin{tabular}{ccccccccc}
					\hline\hline
					\multicolumn{9}{c}{Mean Absolute Error }\\\hline \hline
					\multicolumn{9}{c}{$ Assets=10$ }\\ \hline
					Method&	Sample size	&Min	&1st Qu.	& Median &	Mean	&3rd Qu.&	Max&	Range\\ \hline
					Plug-in &120 &0.1263&	0.2402&	0.3796&	1.1380&	0.7703&	17.6600&	17.5337
					\\ 
					MELO 	&120 &0.0966&	0.2021&	0.2483&	0.2504&	0.2799&	0.4638&	0.3672
					\\
					Plug-in &240 &0.0974&	0.2094&	0.3747&	1.6680&	0.8882&	54.8500&	54.7526
					\\
					MELO 	&240 &0.1012&	0.1966&	0.2160&	0.2320&	0.2606&	0.4689&	0.3677
					\\
					\hline\hline
					\multicolumn{9}{c}{$ Assets=25$ }\\ \hline
					Plug-in &120 & 0.0515&	0.0847&	0.1165&	0.3164&	0.2595&	4.2000&	4.1486
					\\ 
					MELO 	&120 &0.0380&	0.0659&	0.0762&	0.0794&	0.0905&	0.1694&	0.1314
					\\
					Plug-in &240 & 0.0427&	0.0590&	0.0814&	0.1230&	0.1117&	0.8916&	0.8489
					\\
					MELO 	&240 & 0.0380&	0.0547&	0.0653&	0.0683&	0.0785&	0.1575&	0.1195
					\\
					\hline\hline
					\multicolumn{9}{c}{$ Assets=50$ }\\ \hline
					Plug-in &120 &0.0845&	0.1255&	0.1853&	0.8541&	0.3524&	31.1100&	31.0255
					\\ 
					MELO 	&120 & 0.0866&	0.1121&	0.1226&	0.1329&	0.1454&	0.2284&	0.1418
					\\
					Plug-in &240 &0.0633&	0.0997&	0.1233&	0.6502&	0.3045&	8.6480&	8.5847
					\\
					MELO 	&240 &0.0624&	0.0901&	0.0992&	0.1171&	0.1269&	0.2267&	0.1643
					\\
					\hline\hline
					\multicolumn{9}{c}{$ Assets=100$ }\\ \hline
					Plug-in &120 &0.0029&	0.0053&	0.0066&	0.0075&	0.0084&	0.0325&	0.0296
					\\ 
					MELO 	&120 &0.0026&	0.0053&	0.0065&	0.0075&	0.0084&	0.0315&	0.0290
					\\
					Plug-in &240 & 0.0012&	0.0019&	0.0023&	0.0026&	0.0031&	0.0061&	0.0049
					\\
					MELO 	&240 & 0.0012&	0.0019&	0.0023&	0.0026&	0.0031&	0.0060&	0.0049
					\\ \hline\hline
				\end{tabular}
		}}
	\end{table}

	\input{Table4}

	\begin{table}[H]
		\centering
		\caption{Colonial origins of development}\label{origins}
		\begin{tabular}{ccccc}
			\hline\hline
			Method& Column (1)$^1$	& Column (3)$^2$ & Column (5)$^3$& Column (9)$^4$\\
			\hline
			\multirow{2}{*}{MELO (Computational)$^5$} & 0.91 & 1.17	& 0.57 & 0.95\\
			& (0.14) & (0.27) & (0.09) & (0.15) \\
			\hline
			\multirow{2}{*}{2SLS} & 0.94 & 1.28	& 0.58 & 0.98\\
			& (0.16) & (0.36) & (0.10) & (0.17) \\	\hline
			Sample size & 64 & 60 & 37 & 61 \\
			\hline
			$R^2$ First stage & 0.27 & 0.13 & 0.47 & 0.28\\
			\hline
			\hline
			\multicolumn{5}{l}{$^1$ \footnotesize{Base sample.} $^2$ \footnotesize{Base sample without Neo-Europes.} $^3$ \footnotesize{Base sample without Africa.}}\\
			\multicolumn{5}{l}{$^4$ \footnotesize{Base sample, dependent variable is log output per worker.
					Standard error in parentheses.}}\\
			\multicolumn{5}{l}{$^5$ \footnotesize{Using 10,000 iterations.} 
			}
		\end{tabular}	
	\end{table}
\clearpage	
	\section{Appendix}\label{Appendix}
	\subsection{Proof of Proposition 2.1}\label{Proof:proposition1}
	Given $\mathcal{L}({\bf{g}}(\bm{\theta}),\hat{\omega})=({\bf{g}}(\bm{\theta})-\hat{\omega})^T{\bf{Q}}(\bm{\theta})({\bf{g}}(\bm{\theta})-\hat{\omega})$, the posterior expected value of the loss function is
	\begin{equation*}E_{\pi(\bm{\theta}|{\bf{y}})}\left\{\mathcal{L}({\bf{g}}(\bm{\theta}),\hat{\omega})\right\}=E_{\pi(\bm{\theta}|{\bf{y}})}\left\{{\bf{g}}(\bm{\theta})^T{\bf{Q}}(\bm{\theta}){\bf{g}}(\bm{\theta})\right\}-2\hat{\omega}^TE_{\pi(\bm{\theta}|{\bf{y}})}\left\{{\bf{Q}}(\bm{\theta}){\bf{g}}(\bm{\theta})\right\}+\hat{\omega}^TE_{\pi(\bm{\theta}|{\bf{y}})}\left\{{\bf{Q}}(\bm{\theta})\right\}\hat{\omega}
	\end{equation*}
	then
	\begin{equation*}
	\frac{\partial{E_{\pi(\bm{\theta}|{\bf{y}})}\left\{\mathcal{L}({\bf{g}}(\bm{\theta}),\hat{\omega})\right\}}}{\partial{{\hat{\omega}}}}=-2E_{\pi(\bm{\theta}|{\bf{y}})}\left\{{\bf{Q}}(\bm{\theta}){\bf{g}}(\bm{\theta})\right\}+2E_{\pi(\bm{\theta}|{\bf{y}})}\left\{{\bf{Q}}(\bm{\theta})\right\}\hat{\omega}^*={\bf{0}}
	\end{equation*}
	so,
	\begin{align*}
	\hat{\omega}^*({\bf{y}})&=\left[E_{\pi(\bm{\theta}|{\bf{y}})}{\bf{Q}}(\bm{\theta})\right]^{-1}E_{\pi(\bm{\theta}|{\bf{y}})}\left[{\bf{Q}}(\bm{\theta}){\bf{g}}(\bm{\theta})\right]
	\end{align*}
	observe that
	
	\begin{equation*}
	\frac{\partial^2{E_{\pi(\bm{\theta}|{\bf{y}})}\left\{\mathcal{L}({\bf{g}}(\bm{\theta}),\hat{\omega})\right\}}}{\partial{{\hat{\omega}}}\partial{{\hat{\omega}^T}}}=2E_{\pi(\bm{\theta}|{\bf{y}})}\left\{{\bf{Q}}(\bm{\theta})\right\}
	\end{equation*} 
	
	\subsection{Proof of Proposition 2.2}\label{Proof:propositionNew}
	
	\begin{proof}
		We begin by proving the weaker,\footnote{We set $h_k(\bm{\theta})=h(\bm{\theta})$ and $g_k(\bm{\theta})=g(\bm{\theta})$ to simplify notation in this proof.}
		\begin{equation} \label{eq:res1}
		\hat{\mathbf{\omega}}^*\xrightarrow{p}g(\bm{\theta}_0).
		\end{equation}
		
		Setting $\hat{h}=\int h(\bm{\theta})\pi(\bm{\theta}|\bm{y})d\bm{\theta}$, $\bm{u}=\sqrt{N}(\bm{\theta}-\hat{\bm{\theta}})$, then $\pi^*(\bm{u}|\bm{y})=\pi\left(\frac{\bm{u}}{\sqrt{N}}+\hat{\bm{\theta}}|\bm{y}\right)\frac{1}{\sqrt{N}}$. Using a change of variable and taking into account \textbf{F}, we have
		
		\begin{align*}
		|\hat{h}-h(\bm{\theta}_0)|=&\left|\int \left\{h\left(\frac{\bm{u}}{\sqrt{N}}+\hat{\bm{\theta}}\right)-h(\bm{\theta}_0)\right\}\pi\left(\frac{\bm{u}}{\sqrt{N}}+\hat{\bm{\theta}}\left|\bm{y}\right)\frac{1}{\sqrt{N}}d\bm{u}\right|\right.\\
		=&\left|\int \left\{h\left(\frac{\bm{u}}{\sqrt{N}}+\hat{\bm{\theta}}\right)-h(\bm{\theta}_0)\right\}\pi^*(\bm{u}|\bm{y})d\bm{u}\right|\\
		=&\left|\int\left\{ h(\bm{\theta}_0)+\left(\frac{\bm{u}}{\sqrt{N}}+\hat{\bm{\theta}}-\bm{\theta}_0\right)^T\left[\nabla h(\bm{\theta}_0)+V_N\left(\frac{\bm{u}}{\sqrt{N}}+\hat{\bm{\theta}}\right)\right]-h(\bm{\theta}_0)\right\}\pi^*(\bm{u}|\bm{y})d\bm{u}\right|\\
		=&\left|\frac{\nabla h(\bm{\theta}_0)^T}{\sqrt{N}}\int\bm{u}\pi^*(\bm{u}|\bm{y})d\bm{u}+\int V_N\left(\frac{\bm{u}}{\sqrt{N}}+\hat{\bm{\theta}}\right)^T\frac{\bm{u}}{\sqrt{N}}\pi^*(\bm{u}|\bm{y})d\bm{u}\right.\\
		&\left. +(\hat{\bm{\theta}}-\bm{\theta}_0)^T\nabla h(\bm{\theta}_0)+(\hat{\bm{\theta}}-\bm{\theta}_0)^T\int V_N\left(\frac{\bm{u}}{\sqrt{N}}+\hat{\bm{\theta}}\right) \pi^*(\bm{u}|\bm{y})d\bm{u}\right| \\
		\leq&\left|\frac{\nabla h(\bm{\theta}_0)^T}{\sqrt{N}}\int\bm{u}\pi^*(\bm{u}|\bm{y})d\bm{u}\right|+\left|\int sup\left\{\left|V_N\left(\frac{\bm{u}}{\sqrt{N}}+\hat{\bm{\theta}}\right)^T\right|\right\}\frac{\bm{u}}{\sqrt{N}}\pi^*(\bm{u}|\bm{y})d\bm{u}\right|\\
		&+\left|(\hat{\bm{\theta}}-\bm{\theta}_0)^T\nabla h(\bm{\theta}_0)\right|+\left|(\hat{\bm{\theta}}-\bm{\theta}_0)^T\int sup\left\{\left|V_N\left(\frac{\bm{u}}{\sqrt{N}}+\hat{\bm{\theta}}\right)\right|\right\} \pi^*(\bm{u}|\bm{y})d\bm{u}\right|\\
		\leq&\left|\frac{\nabla h(\bm{\theta}_0)^T}{\sqrt{N}}\int\bm{u}(\pi^*(\bm{u}|\bm{y})-\phi(\bm{I}(\bm{\theta}_0)^{-1},\bm{u}))d\bm{u}\right|+\left|\frac{\bm{c}_2^T}{\sqrt{N}}\int \bm{u}(\pi^*(\bm{u}|\bm{y})-\phi(\bm{I}(\bm{\theta}_0)^{-1},\bm{u}))d\bm{u}\right|\\
		&+\left|(\hat{\bm{\theta}}-\bm{\theta}_0)^T\nabla h(\bm{\theta}_0)\right|+\left|(\hat{\bm{\theta}}-\bm{\theta}_0)^T\bm{c}_2\right|\\ 
		\leq&\left\|\frac{\nabla h(\bm{\theta}_0)}{\sqrt{N}}\right\|\int||\bm{u}|| \ |\pi^*(\bm{u}|\bm{y})-\phi(\bm{I}(\bm{\theta}_0)^{-1},\bm{u})|d\bm{u}\\
		&+\frac{c_2}{\sqrt{N}}\int ||\bm{u}|| \ |\pi^*(\bm{u}|\bm{y})-\phi(\bm{I}(\bm{\theta}_0)^{-1},\bm{u})|d\bm{u}+||\hat{\bm{\theta}}-\bm{\theta}_0|| \ ||\nabla h(\bm{\theta}_0)||+c_2||\hat{\bm{\theta}}-\bm{\theta}_0||,\\   
		\end{align*}
		\noindent taking into account assumptions \textbf{F},  $\hat{\bm{\theta}}\xrightarrow{p}\bm{\theta}_0$ and equation (2),
		we conclude that $\hat{h}\xrightarrow{p}h(\bm{\theta}_0)$, that is
		
		\begin{equation}\label{eq:h}
		\int h(\bm{\theta})\pi(\bm{\theta}|\bm{y})d\bm{\theta}\xrightarrow{p}h(\bm{\theta}_0).
		\end{equation}
		
		Now, we prove that	
		
		\begin{equation}\label{eq:gh}
		\int g(\bm{\theta})h(\bm{\theta})\pi(\bm{\theta}|\bm{y})d\bm{\theta}\xrightarrow{p}g(\bm{\theta}_0)h(\bm{\theta}_0).
		\end{equation} 
		
		In particular, setting $\hat{q}=	\int g(\bm{\theta})h(\bm{\theta})\pi(\bm{\theta}|\bm{y})d\bm{\theta}$, 
		
		\begin{align*}
		\left|\hat{q}-g(\bm{\theta}_0)h(\bm{\theta}_0)\right|=&\left|\int \left\{h\left(\frac{\bm{u}}{\sqrt{N}}+\hat{\bm{\theta}}\right)g\left(\frac{\bm{u}}{\sqrt{N}}+\hat{\bm{\theta}}\right)-h(\bm{\theta}_0)g(\bm{\theta}_0)\right\}\pi\left(\frac{\bm{u}}{\sqrt{N}}+\hat{\bm{\theta}}|\bm{y}\right)\frac{1}{\sqrt{N}}d\bm{u}\right|\\
		=&\left|\int \left\{h\left(\frac{\bm{u}}{\sqrt{N}}+\hat{\bm{\theta}}\right)g\left(\frac{\bm{u}}{\sqrt{N}}+\hat{\bm{\theta}}\right)-h(\bm{\theta}_0)g(\bm{\theta}_0)\right\}\pi^*(\bm{u}|\bm{y})d\bm{u}\right|\\
		=&\left|\int\left\{ \left(h(\bm{\theta}_0)+\left(\frac{\bm{u}}{\sqrt{N}}+\hat{\bm{\theta}}-\bm{\theta}_0\right)^T\left[\nabla h(\bm{\theta}_0)+V_N\left(\frac{\bm{u}}{\sqrt{N}}+\hat{\bm{\theta}}\right)\right]\right)\right.\right.\\
		&\left.\left.\left(g(\bm{\theta}_0)+\left(\frac{\bm{u}}{\sqrt{N}}+\hat{\bm{\theta}}-\bm{\theta}_0\right)^T\left[\nabla g(\bm{\theta}_0)+W_N\left(\frac{\bm{u}}{\sqrt{N}}+\hat{\bm{\theta}}\right)\right]\right)-h(\bm{\theta}_0)g(\bm{\theta}_0)\right\} \times \right.\\
		& \left.\pi^*(\bm{u}|\bm{y})d\bm{u}\right|\\
		\leq& h(\bm{\theta}_0)\left|\int\left\{ g\left(\frac{\bm{u}}{\sqrt{N}}+\hat{\bm{\theta}}\right)-g(\bm{\theta}_0)\right\}\pi^*(\bm{u}|\bm{y})d\bm{u}\right|\\
		&+||\hat{\bm{\theta}}-\bm{\theta}_0|| \ ||\nabla h(\bm{\theta}_0)||\left|\int \left\{g\left(\frac{\bm{u}}{\sqrt{N}}+\hat{\bm{\theta}}\right)-g(\bm{\theta}_0)+g(\bm{\theta}_0)\right\}\pi^*(\bm{u}|\bm{y})d\bm{u}\right|\\
		&+||\hat{\bm{\theta}}-\bm{\theta}_0||\bm{c}_2\left|\int \left\{g\left(\frac{\bm{u}}{\sqrt{N}}+\hat{\bm{\theta}}\right)-g(\bm{\theta}_0)+g(\bm{\theta}_0)\right\}\pi^*(\bm{u}|\bm{y})d\bm{u}\right|\\
		&+\left|\frac{1}{\sqrt{N}}g(\bm{\theta}_0)\nabla h(\bm{\theta}_0)^T\int \bm{u}(\pi^*(\bm{u}|\bm{y})-\phi(\bm{I}(\bm{\theta}_0)^{-1},\bm{u}))d\bm{u}\right|\\
		&+\left|\nabla h(\bm{\theta}_0)^T\left(\left[\int \frac{1}{N}\bm{u}\bm{u}^T(\pi^*(\bm{u}|\bm{y})-\phi(\bm{I}(\bm{\theta}_0)^{-1},\bm{u}))d\bm{u}\right]\nabla g(\bm{\theta}_0)+\frac{1}{N}\bm{I}(\bm{\theta}_0)^{-1}\nabla g(\bm{\theta}_0)\right)\right|\\
		&+\left|\nabla h(\bm{\theta}_0)^T\left(\left[\int \frac{1}{N}\bm{u}\bm{u}^T(\pi^*(\bm{u}|\bm{y})-\phi(\bm{I}(\bm{\theta}_0)^{-1},\bm{u}))d\bm{u}\right]\bm{c}_1+\frac{1}{N}\bm{I}(\bm{\theta}_0)^{-1}\bm{c}_1\right)\right|\\
		&+\left|\frac{1}{\sqrt{N}}\nabla h(\bm{\theta}_0)^T\left[\int \bm{u}(\pi^*(\bm{u}|\bm{y})-\phi(\bm{I}(\bm{\theta}_0)^{-1},\bm{u}))d\bm{u}\right](\hat{\bm{\theta}_0}-\bm{\theta}_0)^T\nabla g(\bm{\theta}_0)\right|\\
		&+\left|\frac{1}{\sqrt{N}}\nabla h(\bm{\theta}_0)^T\left[\int \bm{u}(\pi^*(\bm{u}|\bm{y})-\phi(\bm{I}(\bm{\theta}_0)^{-1},\bm{u}))d\bm{u}\right](\hat{\bm{\theta}_0}-\bm{\theta}_0)^T\bm{c}_1\right|\\	
		&+\frac{1}{\sqrt{N}}\left|\bm{c}_2^T\left[\int \bm{u}(\pi^*(\bm{u}|\bm{y})-\phi(\bm{I}(\bm{\theta}_0)^{-1},\bm{u}))d\bm{u}\right]g(\bm{\theta}_0)\right|\\
		&+\left|\bm{c}_2^T\left(\left[\int \frac{1}{N}\bm{u}\bm{u}^T(\pi^*(\bm{u}|\bm{y})-\phi(\bm{I}(\bm{\theta}_0)^{-1},\bm{u}))d\bm{u}\right]\nabla g(\bm{\theta}_0)+\frac{1}{N}\bm{I}(\bm{\theta}_0)^{-1}\nabla g(\bm{\theta}_0)\right)\right|\\
		&+\left|\bm{c}_2^T\left(\left[\int \frac{1}{N}\bm{u}\bm{u}^T(\pi^*(\bm{u}|\bm{y})-\phi(\bm{I}(\bm{\theta}_0)^{-1},\bm{u}))d\bm{u}\right]\nabla g(\bm{\theta}_0)+\frac{1}{N}\bm{I}(\bm{\theta}_0)^{-1}\bm{c}_1\right)\right|\\
		&+\frac{1}{\sqrt{N}}\left|\bm{c}_2^T\left[\int \bm{u}(\pi^*(\bm{u}|\bm{y})-\phi(\bm{I}(\bm{\theta}_0)^{-1},\bm{u}))d\bm{u}\right](\hat{\bm{\theta}}-\bm{\theta}_0)^Tg(\bm{\theta}_0)\right|\\
		&+\frac{1}{\sqrt{N}}\left|\bm{c}_2^T\left[\int \bm{u}(\pi^*(\bm{u}|\bm{y})-\phi(\bm{I}(\bm{\theta}_0)^{-1},\bm{u}))d\bm{u}\right](\hat{\bm{\theta}}-\bm{\theta}_0)^T\bm{c}_1\right|,\\
		\end{align*}
		
		\noindent then,
		
		\begin{align*}
		\left|\hat{q}-g(\bm{\theta}_0)h(\bm{\theta}_0)\right|	\leq&h(\bm{\theta}_0)\left|\int\left\{ g\left(\frac{\bm{u}}{\sqrt{N}}+\hat{\bm{\theta}}\right)-g(\bm{\theta}_0)\right\}\pi^*(\bm{u}|\bm{y})d\bm{u}\right|\\
		&+||\hat{\bm{\theta}}-\bm{\theta}_0|| \ ||\nabla h(\bm{\theta}_0)||\left|\int \left\{g\left(\frac{\bm{u}}{\sqrt{N}}+\hat{\bm{\theta}}\right)-g(\bm{\theta}_0)+g(\bm{\theta}_0)\right\}\pi^*(\bm{u}|\bm{y})d\bm{u}\right|\\
		&+||\hat{\bm{\theta}}-\bm{\theta}_0||\bm{c}_2\left|\int \left\{g\left(\frac{\bm{u}}{\sqrt{N}}+\hat{\bm{\theta}}\right)-g(\bm{\theta}_0)+g(\bm{\theta}_0)\right\}\pi^*(\bm{u}|\bm{y})d\bm{u}\right|\\
		&+\left|\frac{1}{\sqrt{N}}g(\bm{\theta}_0)\nabla h(\bm{\theta}_0)^T\int ||\bm{u}|| \ |\pi^*(\bm{u}|\bm{y})-\phi(\bm{I}(\bm{\theta}_0)^{-1},\bm{u})|d\bm{u}\right|\\
		&+\left|\nabla h(\bm{\theta}_0)^T\left(\left[\int \frac{1}{N}||\bm{u}||^2|\pi^*(\bm{u}|\bm{y})-\phi(\bm{I}(\bm{\theta}_0)^{-1},\bm{u})|d\bm{u}\right]\nabla g(\bm{\theta}_0)+\frac{1}{N}\bm{I}(\bm{\theta}_0)^{-1}\nabla g(\bm{\theta}_0)\right)\right|\\
		&+\left|\nabla h(\bm{\theta}_0)^T\left(\left[\int \frac{1}{N}||\bm{u}||^2|\pi^*(\bm{u}|\bm{y})-\phi(\bm{I}(\bm{\theta}_0)^{-1},\bm{u})|d\bm{u}\right]\bm{c}_1+\frac{1}{N}\bm{I}(\bm{\theta}_0)^{-1}\bm{c}_1\right)\right|\\
		&+\left|\frac{1}{\sqrt{N}}\nabla h(\bm{\theta}_0)^T\left[\int ||\bm{u}|| \ |\pi^*(\bm{u}|\bm{y})-\phi(\bm{I}(\bm{\theta}_0)^{-1},\bm{u})|d\bm{u}\right](\hat{\bm{\theta}_0}-\bm{\theta}_0)^T\nabla g(\bm{\theta}_0)\right|\\
		&+\left|\frac{1}{\sqrt{N}}\nabla h(\bm{\theta}_0)^T\left[\int ||\bm{u}|| \ |\pi^*(\bm{u}|\bm{y})-\phi(\bm{I}(\bm{\theta}_0)^{-1},\bm{u})|d\bm{u}\right](\hat{\bm{\theta}_0}-\bm{\theta}_0)^T\bm{c}_1\right|\\	
		&+\frac{1}{\sqrt{N}}\left|\bm{c}_2^T\left[\int ||\bm{u}|| \ |\pi^*(\bm{u}|\bm{y})-\phi(\bm{I}(\bm{\theta}_0)^{-1},\bm{u})|d\bm{u}\right]g(\bm{\theta}_0)\right|\\
		&+\left|\bm{c}_2^T\left(\left[\int \frac{1}{N}||\bm{u}||^2|\pi^*(\bm{u}|\bm{y})-\phi(\bm{I}(\bm{\theta}_0)^{-1},\bm{u})|d\bm{u}\right]\nabla g(\bm{\theta}_0)+\frac{1}{N}\bm{I}(\bm{\theta}_0)^{-1}\nabla g(\bm{\theta}_0)\right)\right|\\
		&+\left|\bm{c}_2^T\left(\left[\int \frac{1}{N}||\bm{u}||^2|\pi^*(\bm{u}|\bm{y})-\phi(\bm{I}(\bm{\theta}_0)^{-1},\bm{u})|d\bm{u}\right]\nabla g(\bm{\theta}_0)+\frac{1}{N}\bm{I}(\bm{\theta}_0)^{-1}\bm{c}_1\right)\right|\\
		&+\frac{1}{\sqrt{N}}\left|\bm{c}_2^T\left[\int ||\bm{u}|| \ |\pi^*(\bm{u}|\bm{y})-\phi(\bm{I}(\bm{\theta}_0)^{-1},\bm{u})|d\bm{u}\right](\hat{\bm{\theta}}-\bm{\theta}_0)^Tg(\bm{\theta}_0)\right|\\
		&+\frac{1}{\sqrt{N}}\left|\bm{c}_2^T\left[\int ||\bm{u}|| \ |\pi^*(\bm{u}|\bm{y})-\phi(\bm{I}(\bm{\theta}_0)^{-1},\bm{u})|d\bm{u}\right](\hat{\bm{\theta}}-\bm{\theta}_0)^T\bm{c}_1\right|,\\
		\end{align*}

		\noindent taking into account that $	\int g(\bm{\theta})\pi(\bm{\theta}|\bm{y})d\bm{\theta}\xrightarrow{p}g(\bm{\theta}_0)$ by similar arguments that prove \ref{eq:h}, assumptions \textbf{G} and \textbf{F},  $\hat{\bm{\theta}}\xrightarrow{p}\bm{\theta}_0$ and equation (2) 
		we conclude \ref{eq:gh} using algebra as used to prove result \ref{eq:h}.\\
		
		Results \ref{eq:h} and \ref{eq:gh} imply result \ref{eq:res1} by Theorem 2.1.3 in \cite{Lehmann2003} provided $h(\bm{\theta}_0)\neq 0$.\\
		
		Now,
		
		\begin{align*}
		\sqrt{N}(\hat{\mathbf{\omega}}^*-g(\bm{\theta}_0))=&\sqrt{N}(\hat{\mathbf{\omega}}^*-g(\hat{\bm{\theta}}))+\sqrt{N}(g(\hat{\bm{\theta}})-g(\bm{\theta}_0)),
		\end{align*}
		\noindent given that $\sqrt{N}(\hat{\bm{\theta}}-\bm{\theta}_0)\xrightarrow{d}\mathcal{N}(\bm{0},[I(\bm{\theta}_0)]^{-1})$, then $\sqrt{N}(g(\hat{\bm{\theta}})-g(\bm{\theta}_0))\xrightarrow{d}\mathcal{N}(\bm{0},\nabla g(\bm{\theta}_0)^T[I(\bm{\theta}_0)]^{-1}g(\bm{\theta}_0))$ by the delta method. So, it only remains to show that $\sqrt{N}(\hat{\mathbf{\omega}}^*-g(\hat{\bm{\theta}}))\xrightarrow{p}0$ by the Slutsky's theorem.\\
		
		Making a Taylor expansion for $h(\bm{\theta})$ at $\bm{\theta}_0$, and for $g(\bm{\theta})$ at $\hat{\bm{\theta}}$,  	
		\begin{align*}
		\sqrt{N}(\hat{\mathbf{\omega}}^*-g(\hat{\bm{\theta}}))=&\frac{1}{\hat{h}}\left\{h(\bm{\theta}_0)\int \bm{u}^T\left[\nabla g(\hat{\bm{\theta}})+W_N\left(\frac{\bm{u}}{\sqrt{N}}+\hat{\bm{\theta}}\right)\right]\pi^*(\bm{u}|\bm{y})d\bm{u}+\right.\\
		&\left.\frac{1}{\sqrt{N}}\int (\bm{u}+\hat{\bm{\theta}}-\bm{\theta}_0)^T\left[\nabla h(\bm{\theta}_0)+V_N\left(\frac{\bm{u}}{\sqrt{N}}+\hat{\bm{\theta}}\right)\right] \bm{u}^T \times \right.\\
		& \left.\left[\nabla g(\hat{\bm{\theta}})+W_N\left(\frac{\bm{u}}{\sqrt{N}}+\hat{\bm{\theta}}\right)\right]\pi^*(\bm{u}|\bm{y})d\bm{u}\right\}.\\
		\leq&\frac{1}{\hat{h}}\left\{h(\bm{\theta}_0)\left|\left\{\int \bm{u}^T(\pi^*(\bm{u}|\bm{y})-\phi(\bm{I}(\bm{\theta}_0)^{-1},\bm{u}))d\bm{u}\right\}\nabla g(\hat{\bm{\theta}})\right|\right.\\
		&+h(\bm{\theta}_0)\left|\left\{\int \bm{u}^T(\pi^*(\bm{u}|\bm{y})-\phi(\bm{I}(\bm{\theta}_0)^{-1},\bm{u}))d\bm{u}\right\}\bm{c}_1\right|\\
		&+\frac{1}{\sqrt{N}}\left|\nabla h(\bm{\theta}_0)^T\left(\int \bm{u}\bm{u}^T(\pi^*(\bm{u}|\bm{y})-\phi(\bm{I}(\bm{\theta}_0)^{-1},\bm{u}))d\bm{u}+\bm{I}(\bm{\theta}_0)^{-1}\right)\nabla g(\hat{\bm{\theta}})\right|\\
		&+\frac{1}{\sqrt{N}}\left|\bm{c}_2^T\left(\int \bm{u}\bm{u}^T(\pi^*(\bm{u}|\bm{y})-\phi(\bm{I}(\bm{\theta}_0)^{-1},\bm{u}))d\bm{u}+\bm{I}(\bm{\theta}_0)^{-1}\right)\nabla g(\hat{\bm{\theta}})\right|\\
		&+\frac{1}{\sqrt{N}}\left|\nabla h(\bm{\theta}_0)^T\left(\int \bm{u}\bm{u}^T(\pi^*(\bm{u}|\bm{y})-\phi(\bm{I}(\bm{\theta}_0)^{-1},\bm{u}))d\bm{u}+\bm{I}(\bm{\theta}_0)^{-1}\right)\bm{c}_1\right|\\
		&+\frac{1}{\sqrt{N}}\left|\bm{c}_2^T\left(\int \bm{u}\bm{u}^T(\pi^*(\bm{u}|\bm{y})-\phi(\bm{I}(\bm{\theta}_0)^{-1},\bm{u}))d\bm{u}+\bm{I}(\bm{\theta}_0)^{-1}\right)\bm{c}_1\right|\\
		&+\frac{1}{\sqrt{N}}\left|(\hat{\bm{\theta}}-\bm{\theta}_0)^T\nabla h(\bm{\theta}_0)\int\left\{ \bm{u}^T(\pi^*(\bm{u}|\bm{y})-\phi(\bm{I}(\bm{\theta}_0)^{-1},\bm{u}))d\bm{u}\right\}\nabla g(\hat{\bm{\theta}})\right|\\
		&+\frac{1}{\sqrt{N}}\left|(\hat{\bm{\theta}}-\bm{\theta}_0)^T\bm{c}_2\int\left\{ \bm{u}^T(\pi^*(\bm{u}|\bm{y})-\phi(\bm{I}(\bm{\theta}_0)^{-1},\bm{u}))d\bm{u}\right\}\nabla g(\hat{\bm{\theta}})\right|\\
		&+\frac{1}{\sqrt{N}}\left|(\hat{\bm{\theta}}-\bm{\theta}_0)^T\nabla h(\bm{\theta}_0)\int\left\{ \bm{u}^T(\pi^*(\bm{u}|\bm{y})-\phi(\bm{I}(\bm{\theta}_0)^{-1},\bm{u}))d\bm{u}\right\}\bm{c}_1\right|\\
		&\left.+\frac{1}{\sqrt{N}}\left|(\hat{\bm{\theta}}-\bm{\theta}_0)^T\bm{c}_2\int\left\{ \bm{u}^T(\pi^*(\bm{u}|\bm{y})-\phi(\bm{I}(\bm{\theta}_0)^{-1},\bm{u}))d\bm{u}\right\}\bm{c}_1\right|\right\},\\
		\end{align*}	
		\noindent taking into account assumptions \textbf{G} and \textbf{F},  $\hat{\bm{\theta}}\xrightarrow{p}\bm{\theta}_0$ and equation (2) 
		we have by result \ref{eq:h} that $\hat{h}\xrightarrow{p}h(\bm{\theta}_0)\neq 0$, and using algebra as used to prove result \ref{eq:h}, the numerator converges in probability to 0, then  we conclude that $\sqrt{N}(\hat{\mathbf{\omega}}^*-g(\hat{\bm{\theta}}))\xrightarrow{p}0$ by Theorem 2.1.3 in \cite{Lehmann2003}.\\
	\end{proof}
	
	\subsection{Proof of Proposition 2.3}\label{Proof:proposition2}
	\begin{align}
	\hat{\omega}^*({\bf{y}})&=\left[\int_{\Theta}{\bf{Q}}(\bm{\theta})\pi(\bm{\theta}|{\bf{y}})d\bm{\theta}\right]^{-1}\left[\int_{\Theta}{\bf{Q}}(\bm{\theta}){\bf{g}}(\bm{\theta})\pi(\bm{\theta}|{\bf{y}})d\bm{\theta}\right]\nonumber
	\end{align}
	\noindent where 
	\begin{align*}
	\pi(\bm{\theta}|{\bf{y}}) &=\frac{\pi(\bm{\theta})f({\bf{y}}|\bm{\theta})}{\int{\pi(\bm{\theta})f({\bf{y}}|\bm{\theta})d\bm{\theta}}}\\
	&=\frac{\pi(\bm{\theta})w({\bf{y}})h(\hat{\bm{\theta}}|\bm{\theta})}{\int{\pi(\bm{\theta})w({\bf{y}})h(\hat{\bm{\theta}}|\bm{\theta})d\bm{\theta}}}
	\end{align*}
	\noindent The second equality follows from the Factorization theorem due to $\hat{\bm{\theta}}$ is a sufficient statistic, $w({\bf{y}})$ does not depend on $\bm{\theta}$, and $h(\hat{\bm{\theta}}|\bm{\theta})$ depends on ${\bf{y}}$ only through $\hat{\bm{\theta}}$.
	Then,\\
	\begin{align}
	\hat{\omega}^*({\bf{y}})&=\left[\int_{\Theta}{\bf{Q}}(\bm{\theta})\frac{\pi(\bm{\theta})w({\bf{y}})h(\hat{\bm{\theta}}|\bm{\theta})}{\int{\pi(\bm{\theta})w({\bf{y}})h(\hat{\bm{\theta}}|\bm{\theta})d\bm{\theta}}}d\bm{\theta}\right]^{-1}\left[\int_{\Theta}{\bf{Q}}(\bm{\theta}){\bf{g}}(\bm{\theta})\frac{\pi(\bm{\theta})w({\bf{y}})h(\hat{\bm{\theta}}|\bm{\theta})}{\int{\pi(\bm{\theta})w({\bf{y}})h(\hat{\bm{\theta}}|\bm{\theta})d\bm{\theta}}}d\bm{\theta}\right]\nonumber\\
	&=\left[\int_{\Theta}{\bf{Q}}(\bm{\theta})\pi(\bm{\theta})h(\hat{\bm{\theta}}|\bm{\theta})d\bm{\theta}\right]^{-1}\left[\int_{\Theta}{\bf{Q}}(\bm{\theta}){\bf{g}}(\bm{\theta})\pi(\bm{\theta})h(\hat{\bm{\theta}}|\bm{\theta})d\bm{\theta}\right]\nonumber\\
	&=\left[\int_{\Theta}{\bf{Q}}(\bm{\theta})\pi(\bm{\theta}|\hat{\bm{\theta}})d\bm{\theta}\right]^{-1}\left[\int_{\Theta}{\bf{Q}}(\bm{\theta}){\bf{g}}(\bm{\theta})\pi(\bm{\theta}|\hat{\bm{\theta}})d\bm{\theta}\right]\nonumber\\
	&=\left[E_{\pi(\bm{\theta}|\hat{\bm{\theta}}({\bf{y}}))}{\bf{Q}}(\bm{\theta})\right]^{-1}E_{\pi(\bm{\theta}|\hat{\bm{\theta}}({\bf{y}}))}\left[{\bf{Q}}(\bm{\theta}){\bf{g}}(\bm{\theta})\right]\nonumber\\
	&=\hat{\omega}^*(\hat{\bm{\theta}}({\bf{y}}))\nonumber
	\end{align}
	
	\noindent The third equality is due to the likelihood principle, that is, all the relevant information regarding $\bm{\theta}$ is in $h(\hat{\bm{\theta}}|\bm{\theta})$ \citep{berger93,Bernardo1994}.

	\subsection{Proof of Lemma 2.4}\label{Proof:lemma1}
	
	\begin{align*}
	\hat{\omega}^*({\bf{y}}) = \hat{\omega}^*(\hat{\bm{\theta}}({\bf{y}})) & = [{E_{\pi(\bm{\theta}|\hat{\bm{\theta}}({\bf{y}}))}{\bf{Q}}(\bm{\theta})]}^{-1} E_{\pi(\bm{\theta}|\hat{\bm{\theta}}({\bf{y}}))}[{\bf{Q}}(\bm{\theta}) {\bf{g}}(\bm{\theta})] \\
	& = \left\{\int {\bf{Q}}(\bm{\theta}) \pi(\bm{\theta}) h(\hat{\bm{\theta}}|\bm{\theta}) d\bm{\theta}\right\}^{-1} \left\{\int {\bf{Q}}(\bm{\theta}) {\bf{g}}(\bm{\theta}) \pi(\bm{\theta}) h(\hat{\bm{\theta}}|\bm{\theta}) d\bm{\theta}\right\}
	\end{align*}
	Denoting 
	\begin{align*}
	A & = \int {\bf{Q}}(\bm{\theta}) \pi(\bm{\theta}) h(\hat{\bm{\theta}}|\bm{\theta}) d\bm{\theta} \\
	B & = \int {\bf{Q}}(\bm{\theta}) {\bf{g}}(\bm{\theta}) \pi(\bm{\theta}) h(\hat{\bm{\theta}}|\bm{\theta}) d\bm{\theta}
	\end{align*}
	
	of order $K\times K$ and $K\times 1$ respectively.
	So, $\hat{\omega} = A^{-1}B$, applying the properties of the differentiation of matrices,\footnote{If $A$ and $B$ are matrices of order $N\times K$ and $K\times M$, and $x$ is a vector of order $1 \times N$, then $\nabla_{x}AB = \nabla_{x}A[B \otimes I_{N}] + A \nabla_{x} B$.}
	
	\begin{equation*}
	\nabla_{\hat{\bm{\theta}}({\bf{y}})}\hat{\omega}^*(\hat{\bm{\theta}}({\bf{y}})) =  A^{-1} \nabla_{\hat{\bm{\theta}}({\bf{y}})}B + \nabla_{\hat{\bm{\theta}}({\bf{y}})} A^{-1} \left[B \otimes I_{P}\right]
	\end{equation*}
	
	\noindent where 
	\begin{equation*}
	\nabla_{\hat{\bm{\theta}}({\bf{y}})} A^{-1} = - A^{-1} \nabla_{\hat{\bm{\theta}}({\bf{y}})} A \left[A^{-1} \otimes I_{N}\right] \footnote{$0 = \nabla_{\hat{\bm{\theta}}({\bf{y}})}AA^{-1} = \nabla_{\hat{\bm{\theta}}({\bf{y}})}A[ A^{-1}\otimes I_{P}] + A^{-1} \nabla_{\hat{\bm{\theta}}({\bf{y}})} A^{-1}$, then $\nabla A^{-1} = - A^{-1} \nabla_{\hat{\bm{\theta}}({\bf{y}})} A \left[A^{-1} \otimes I_{P}\right]$.}
	\end{equation*}
	
	Therefore,\footnote{For the outcome of the third equality, let $A_{1}$, $A_{2}$, $B_{1}$ and $B_{2}$ be matrices of orders $M \times N$, $M \times P$, $L \times R$ and $R \times P$, then $\left(A_{1} \otimes B_{1}\right)\left(A_{2} \otimes B_{2}\right) = A_{1}A_{2} \otimes B_{1}B_{2}$.}
	\begin{align*}
	\nabla_{\hat{\bm{\theta}}({\bf{y}})}\hat{\omega}^*(\hat{\bm{\theta}}({\bf{y}})) & = A^{-1} \nabla_{\hat{\bm{\theta}}({\bf{y}})} B - A^{-1} \nabla_{\hat{\bm{\theta}}({\bf{y}})} A \left[A^{-1} \otimes I_{P}\right]\left[B \otimes I_{P}\right] \\
	& =  A^{-1} \left\{\nabla_{\hat{\bm{\theta}}({\bf{y}})} B -  \nabla_{\hat{\bm{\theta}}({\bf{y}})} A \left[A^{-1} \otimes I_{P}\right]\left[B \otimes I_{P}\right]\right\} \\
	& =  A^{-1} \left\{\nabla_{\hat{\bm{\theta}}({\bf{y}})} B -  \nabla_{\hat{\bm{\theta}}({\bf{y}})} A \left[A^{-1} B \otimes I_{P}\right]\right\}  \\
	& =  A^{-1} \left\{\nabla_{\hat{\bm{\theta}}({\bf{y}})} B -  \nabla_{\hat{\bm{\theta}}({\bf{y}})} A \left[\hat{\omega} \otimes I_{P}\right]\right\} 
	\end{align*}
	
	\noindent where
	\noindent $\nabla_{\hat{\bm{\theta}}({\bf{y}})} \hat{\omega}^*({\bf{y}})$ is a $K\times P$ matrix of partial derivatives, that is,
	\begin{equation*}
	\nabla_{\hat{\bm{\theta}}({\bf{y}})} \hat{\omega}^*(\hat{\bm{\theta}}({\bf{y}}))= \begin{bmatrix}
	\frac{\partial{\hat{\omega}_1^*}}{\partial{\hat{\bm{\theta}}({\bf{y}})_1}}&  \frac{\partial{\hat{\omega}_1^*}}{\partial{\hat{\bm{\theta}}({\bf{y}})_2}} & \dots &  \frac{\partial{\hat{\omega}_1^*}}{\partial{\hat{\bm{\theta}}({\bf{y}})_P}}  \\ 
	\frac{\partial{\hat{\omega}_2^*}}{\partial{\hat{\bm{\theta}}({\bf{y}})_1}}&  \frac{\partial{\hat{\omega}_2^*}}{\partial{\hat{\bm{\theta}}({\bf{y}})_2}} & \dots &  \frac{\partial{\hat{\omega}_2^*}}{\partial{\hat{\bm{\theta}}({\bf{y}})_P}}  \\
	\vdots & \vdots & \ddots & \vdots \\
	\frac{\partial{\hat{\omega}_K^*}}{\partial{\hat{\bm{\theta}}({\bf{y}})_1}}&  \frac{\partial{\hat{\omega}_K^*}}{\partial{\hat{\bm{\theta}}({\bf{y}})_2}} & \dots &  \frac{\partial{\hat{\omega}_K^*}}{\partial{\hat{\bm{\theta}}({\bf{y}})_P}}   
	\end{bmatrix},
	\end{equation*}
	\noindent  
	\begin{align*}
	\nabla_{\hat{\bm{\theta}}({\bf{y}})} A & = \int {\bf{Q}}(\bm{\theta}) \pi(\bm{\theta}) \nabla_{\hat{\bm{\theta}}({\bf{y}})} h(\hat{\bm{\theta}}|\bm{\theta}) d\bm{\theta}\\
	& = \int ({\bf{Q}}(\bm{\theta})\otimes \alpha_{\hat{\bm{\theta}}({\bf{y}})}(\bm{\theta})) \pi(\bm{\theta})  h(\hat{\bm{\theta}}|\bm{\theta}) d\bm{\theta} \\
	\end{align*}
	\noindent and
	
	\begin{align*}
	\nabla_{\hat{\bm{\theta}}({\bf{y}})} B & = \int {\bf{Q}}(\bm{\theta}) {\bf{g}}(\bm{\theta}) \pi(\bm{\theta}) \nabla_{\hat{\bm{\theta}}({\bf{y}})} h(\hat{\bm{\theta}}|\bm{\theta}) d\bm{\theta}\\
	& = \int (({\bf{Q}}(\bm{\theta}) {\bf{g}}(\bm{\theta}))\otimes \alpha_{\hat{\bm{\theta}}({\bf{y}})}(\bm{\theta})) \pi(\bm{\theta})  h(\hat{\bm{\theta}}|\bm{\theta}) d\bm{\theta}\\
	\end{align*}
	
	\noindent so,
	
	{\scriptsize{
			\begin{align*}
			\nabla_{\hat{\bm{\theta}}({\bf{y}})}\hat{\omega}^*(\hat{\bm{\theta}}({\bf{y}}))  = & \left\{\int {\bf{Q}}(\bm{\theta}) \pi(\bm{\theta}) h(\hat{\bm{\theta}}|\bm{\theta}) d\bm{\theta}\right\}^{-1} \\ 
			\times &  \left\{\int (({\bf{Q}}(\bm{\theta}) {\bf{g}}(\bm{\theta}))\otimes \alpha_{\hat{\bm{\theta}}({\bf{y}})}(\bm{\theta})) \pi(\bm{\theta})  h(\hat{\bm{\theta}}|\bm{\theta}) d\bm{\theta} - \left[\int ({\bf{Q}}(\bm{\theta})\otimes \alpha_{\hat{\bm{\theta}}({\bf{y}})}(\bm{\theta})) \pi(\bm{\theta})  h(\hat{\bm{\theta}}|\bm{\theta}) d\bm{\theta}\right] \left[\hat{\omega} \otimes I_{P}\right]\right\} \nonumber \\
			= & \left\{E_{\pi(\hat{\bm{\theta}}({\bf{y}}))}[Q(\bm{\theta})]\right\}^{-1} \left\{E_{\pi(\hat{\bm{\theta}}({\bf{y}}))}[(Q(\bm{\theta})g(\bm{\theta}))\otimes\alpha_{\hat{\bm{\theta}}({\bf{y}})}(\bm{\theta})] - E_{\pi(\hat{\bm{\theta}}({\bf{y}}))}[Q(\bm{\theta})\otimes\alpha_{\hat{\bm{\theta}}({\bf{y}})}(\bm{\theta})]\left[\hat{\omega} \otimes I_{P}\right]\right\}
			\end{align*}
		}}

		\noindent where

		\begin{align*}
		& E[(Q(\bm{\theta})g(\bm{\theta}))\otimes \alpha_{\hat{\bm{\theta}}({\bf{y}})}(\bm{\theta})] = \\
		& \begin{bmatrix}
		E\sum_{j=1}^KQ_{1j}g_j\frac{\partial{log h(\hat{\bm{\theta}}|\bm{\theta})}}{\partial{\hat{\bm{\theta}}({\bf{y}})_1}}&  E\sum_{j=1}^KQ_{1j}g_j\frac{\partial{log h(\hat{\bm{\theta}}|\bm{\theta})}}{\partial{\hat{\bm{\theta}}({\bf{y}})_2}} & \dots &E  \sum_{j=1}^KQ_{1j}g_j\frac{\partial{log h(\hat{\bm{\theta}}|\bm{\theta})}}{\partial{\hat{\bm{\theta}}({\bf{y}})_P}}  \\ 
		E\sum_{j=1}^KQ_{2j}g_j\frac{\partial{log h(\hat{\bm{\theta}}|\bm{\theta})}}{\partial{\hat{\bm{\theta}}({\bf{y}})_1}}&  E\sum_{j=1}^KQ_{2j}g_j\frac{\partial{log h(\hat{\bm{\theta}}|\bm{\theta})}}{\partial{\hat{\bm{\theta}}({\bf{y}})_2}} & \dots &E\sum_{j=1}^KQ_{2j}g_j\frac{\partial{log h(\hat{\bm{\theta}}|\bm{\theta})}}{\partial{\hat{\bm{\theta}}({\bf{y}})_P}}  \\
		\vdots & \vdots & \ddots & \vdots \\
		E\sum_{j=1}^KQ_{Kj}g_j\frac{\partial{log h(\hat{\bm{\theta}}|\bm{\theta})}}{\partial{\hat{\bm{\theta}}({\bf{y}})_1}}&  E\sum_{j=1}^KQ_{Kj}g_j\frac{\partial{log h(\hat{\bm{\theta}}|\bm{\theta})}}{\partial{\hat{\bm{\theta}}({\bf{y}})_2}} & \dots &E\sum_{j=1}^KQ_{Kj}g_j\frac{\partial{log h(\hat{\bm{\theta}}|\bm{\theta})}}{\partial{\hat{\bm{\theta}}({\bf{y}})_P}}  \\  
		\end{bmatrix}
		\end{align*}
		
		\noindent and
		{\scriptsize{
				\begin{align*}
				&E[Q(\bm{\theta})\otimes\alpha_{\hat{\bm{\theta}}({\bf{y}})}(\bm{\theta})] = \\
				& {\tiny{ 
						\begin{bmatrix}
						EQ_{11}\frac{\partial{log h(\hat{\bm{\theta}}|\bm{\theta})}}{\partial{\hat{\bm{\theta}}({\bf{y}})_1}}&  EQ_{11}\frac{\partial{log h(\hat{\bm{\theta}}|\bm{\theta})}}{\partial{\hat{\bm{\theta}}({\bf{y}})_2}} & \dots &EQ_{11}\frac{\partial{log h(\hat{\bm{\theta}}|\bm{\theta})}}{\partial{\hat{\bm{\theta}}({\bf{y}})_P}} &EQ_{12}\frac{\partial{log h(\hat{\bm{\theta}}|\bm{\theta})}}{\partial{\hat{\bm{\theta}}({\bf{y}})_1}}&EQ_{12}\frac{\partial{log h(\hat{\bm{\theta}}|\bm{\theta})}}{\partial{\hat{\bm{\theta}}({\bf{y}})_2}}&\dots&EQ_{1K}\frac{\partial{log h(\hat{\bm{\theta}}|\bm{\theta})}}{\partial{\hat{\bm{\theta}}({\bf{y}})_P}}  \\ EQ_{21}\frac{\partial{log h(\hat{\bm{\theta}}|\bm{\theta})}}{\partial{\hat{\bm{\theta}}({\bf{y}})_1}}&  EQ_{21}\frac{\partial{log h(\hat{\bm{\theta}}|\bm{\theta})}}{\partial{\hat{\bm{\theta}}({\bf{y}})_2}} & \dots &EQ_{21}\frac{\partial{log h(\hat{\bm{\theta}}|\bm{\theta})}}{\partial{\hat{\bm{\theta}}({\bf{y}})_P}} &EQ_{22}\frac{\partial{log h(\hat{\bm{\theta}}|\bm{\theta})}}{\partial{\hat{\bm{\theta}}({\bf{y}})_1}}&EQ_{22}\frac{\partial{log h(\hat{\bm{\theta}}|\bm{\theta})}}{\partial{\hat{\bm{\theta}}({\bf{y}})_2}}&\dots&EQ_{2K}\frac{\partial{log h(\hat{\bm{\theta}}|\bm{\theta})}}{\partial{\hat{\bm{\theta}}({\bf{y}})_P}}  \\ 
						\vdots & \vdots & \ddots & \vdots & \vdots & \vdots & \ddots & \vdots \\
						EQ_{K1}\frac{\partial{log h(\hat{\bm{\theta}}|\bm{\theta})}}{\partial{\hat{\bm{\theta}}({\bf{y}})_1}}&  EQ_{K1}\frac{\partial{log h(\hat{\bm{\theta}}|\bm{\theta})}}{\partial{\hat{\bm{\theta}}({\bf{y}})_2}} & \dots &EQ_{K1}\frac{\partial{log h(\hat{\bm{\theta}}|\bm{\theta})}}{\partial{\hat{\bm{\theta}}({\bf{y}})_P}} &EQ_{K2}\frac{\partial{log h(\hat{\bm{\theta}}|\bm{\theta})}}{\partial{\hat{\bm{\theta}}({\bf{y}})_1}}&EQ_{K2}\frac{\partial{log h(\hat{\bm{\theta}}|\bm{\theta})}}{\partial{\hat{\bm{\theta}}({\bf{y}})_2}}&\dots&EQ_{KK}\frac{\partial{log h(\hat{\bm{\theta}}|\bm{\theta})}}{\partial{\hat{\bm{\theta}}({\bf{y}})_P}}  \\
						\end{bmatrix}}}
				\end{align*}
			}}
			
			\subsection{Proof Corollary 2.5}\label{Proof:corollary1}
			\begin{align*}
			\hat{\omega}^*(\hat{\bm{\theta}}({\bf{y}})) & = \frac{E_{\pi(\hat{\bm{\theta}}({\bf{y}}))}[Q(\bm{\theta})g(\bm{\theta})|\hat{\bm{\theta}}({\bf{y}})]}{E_{\pi(\hat{\bm{\theta}}({\bf{y}}))}[Q(\bm{\theta})|\hat{\bm{\theta}}({\bf{y}})]} \\
			& = \frac{\left[\displaystyle\int Q(\bm{\theta}) g(\bm{\theta}) \pi(\bm{\theta}) h(\hat{\bm{\theta}}|\bm{\theta}) d\bm{\theta}\right]/c}{\left[\displaystyle\int Q(\bm{\theta}) \pi(\bm{\theta}) h(\hat{\bm{\theta}}|\bm{\theta}) d\bm{\theta}\right]/c}
			\end{align*}
			
			\noindent 
			where $$c = \displaystyle\int \pi(\bm{\theta}) h(\hat{\bm{\theta}}|\bm{\theta}) d\bm{\theta}.$$

			Let \begin{align*}
			A & = \int Q(\bm{\theta}) g(\bm{\theta}) \pi(\bm{\theta}) h(\hat{\bm{\theta}}|\bm{\theta}) d\bm{\theta}  \\
			B & = \int Q(\bm{\theta}) \pi(\bm{\theta}) h(\hat{\bm{\theta}}|\bm{\theta}) d\bm{\theta}.
			\end{align*}
			
			\noindent then
			
			\begin{align*}
			A' & = \int Q(\bm{\theta}) g(\bm{\theta}) \alpha_{\hat{\bm{\theta}}({\bf{y}})}(\bm{\theta}) \pi(\bm{\theta}) h(\hat{\bm{\theta}}|\bm{\theta}) d\bm{\theta} \\
			B' & = \int Q(\bm{\theta}) \alpha_{\hat{\bm{\theta}}({\bf{y}})}(\bm{\theta}) \pi(\bm{\theta}) h(\hat{\bm{\theta}}|\bm{\theta}) d\bm{\theta}
			\end{align*}
			
			Therefore, 
			
			\begin{align*}
			\nabla_{\hat{\bm{\theta}}(\bm{y})}\hat{\omega}^*(\hat{\bm{\theta}}({\bf{y}}))   = & \frac{A}{B} \left\{\frac{A'}{A} - \frac{B'}{B}\right\} \\
			= & \frac{E_{\pi(\bm{\theta}|\hat{\bm{\theta}}({\bf{y}}))}[Q(\bm{\theta}) g(\bm{\theta})|\hat{\bm{\theta}}({\bf{y}})]}{E_{\pi(\bm{\theta}|\hat{\bm{\theta}}({\bf{y}}))}[Q(\bm{\theta})|\hat{\bm{\theta}}({\bf{y}})]} \\
			\times &  \left\{\frac{E_{\pi(\bm{\theta}|\hat{\bm{\theta}}({\bf{y}}))}[Q(\bm{\theta}) g(\bm{\theta}) \alpha_{\hat{\bm{\theta}}({\bf{y}})}(\bm{\theta})|\hat{\bm{\theta}}({\bf{y}})]}{E_{\pi(\bm{\theta}|\hat{\bm{\theta}}({\bf{y}}))}[Q(\bm{\theta}) g(\bm{\theta})|\hat{\bm{\theta}}({\bf{y}})]}  - \frac{E_{\pi(\bm{\theta}|\hat{\bm{\theta}}({\bf{y}}))}[Q(\bm{\theta}) \alpha_{\hat{\bm{\theta}}({\bf{y}})}(\bm{\theta})|\hat{\bm{\theta}}({\bf{y}})]}{E_{\pi(\bm{\theta}|\hat{\bm{\theta}}({\bf{y}}))}[Q(\bm{\theta})|\hat{\bm{\theta}}({\bf{y}})]}\right\} \\
			= & \frac{E_{\pi(\bm{\theta}|\hat{\bm{\theta}}({\bf{y}}))}[Q(\bm{\theta}) g(\bm{\theta}) \alpha_{\hat{\bm{\theta}}({\bf{y}})}(\bm{\theta})|\hat{\bm{\theta}}({\bf{y}})]}{E_{\pi(\bm{\theta}|\hat{\bm{\theta}}({\bf{y}}))}[Q(\bm{\theta})|\hat{\bm{\theta}}({\bf{y}})]} \\
			- &  \frac{E_{\pi(\bm{\theta}|\hat{\bm{\theta}}({\bf{y}}))}[Q(\bm{\theta}) g(\bm{\theta})|\hat{\bm{\theta}}({\bf{y}})] E_{\pi(\bm{\theta}|\hat{\bm{\theta}}({\bf{y}}))}[Q(\bm{\theta}) \alpha_{\hat{\bm{\theta}}({\bf{y}})}(\bm{\theta})|\hat{\bm{\theta}}({\bf{y}})]}{(E_{\pi(\bm{\theta}|\hat{\bm{\theta}}({\bf{y}}))}[Q(\bm{\theta})|\hat{\bm{\theta}}({\bf{y}})])^2}
			\end{align*}
			
			\subsection{Proof of Proposition 2.6}\label{Proof:proposition3}
			Given a sufficient statistic such that $\hat{\bm{\theta}}({\bf{y}})\sim(\mu_{\bm{\theta}},\Sigma_{\bm{\theta}})$, then a second order Taylor expansion implies that 
			\begin{equation*}
			\hat{\omega}^*({\bf{y}})= \hat{\omega}^*(\hat{\bm{\theta}}({\bf{y}}))\approx \hat{\omega}^*(\bm{\theta}_0)+\nabla_{\bm{\theta}}\hat{\omega}^*(\bm{\theta}_0)(\hat{\bm{\theta}}-\bm{\theta}_0)
			\end{equation*}
			\noindent Then the variance of $\hat{\omega}^*(\hat{\bm{\theta}}({\bf{y}}))$ is 
			
			\begin{equation*}
			Var(\hat{\omega}^*({\bf{y}}))=Var(\hat{\omega}^*(\hat{\bm{\theta}}({\bf{y}})))\approx \nabla_{\bm{\theta}}\hat{\omega}^*(\bm{\theta}_0) \Sigma_{\hat{\bm{\theta}}} \nabla_{\bm{\theta}}\hat{\omega}^*(\bm{\theta}_0)^T
			\end{equation*}
			
			\noindent if $\hat{\bm{\theta}}\xrightarrow{p}\bm{\theta}_0$, then 
			
			\begin{equation*}
			Var(\hat{\omega}^*({\bf{y}}))=Var(\hat{\omega}^*(\bm{\theta}({\bf{y}})))\approx \nabla_{\bm{\theta}}\hat{\omega}^*(\hat{\bm{\theta}}) \Sigma_{\hat{\bm{\theta}}} \nabla_{\bm{\theta}}\hat{\omega}^*(\hat{\bm{\theta}})^T
			\end{equation*}     
			 
\end{document}

%% file: Table2.tex
\begin{sidewaystable}
  \centering
  \caption{Odds ratio problem: Mean Errors for $x=(1,1,1)$}
	{\scriptsize
    \begin{tabular}{rrrrrrrrr}
    \hline
    \hline
    \multicolumn{9}{c}{Mean Square Error} \\
    \hline
    \hline
    \multicolumn{1}{c}{Method} & \multicolumn{1}{c}{Sample size} & \multicolumn{1}{c}{Min} & \multicolumn{1}{c}{1st Qu.} & \multicolumn{1}{c}{Median} & \multicolumn{1}{c}{Mean} & \multicolumn{1}{c}{3rd Qu.} & \multicolumn{1}{c}{Max} & \multicolumn{1}{c}{Range} \\
    \hline
    \multicolumn{1}{c}{Plug-in*} & \multicolumn{1}{c}{20} & \multicolumn{1}{c}{0.0000} & \multicolumn{1}{c}{0.2484} & \multicolumn{1}{c}{0.7826} & \multicolumn{1}{c}{3.7586E+17} & \multicolumn{1}{c}{1.8843} & \multicolumn{1}{c}{1.7336E+20} & \multicolumn{1}{c}{1.7336E+20} \\
    \multicolumn{1}{c}{MELO**} & \multicolumn{1}{c}{20} & \multicolumn{1}{c}{0.0000} & \multicolumn{1}{c}{0.1683} & \multicolumn{1}{c}{0.5962} & \multicolumn{1}{c}{0.9860} & \multicolumn{1}{c}{1.1016} & \multicolumn{1}{c}{29.7273} & \multicolumn{1}{c}{29.7273} \\
    \multicolumn{1}{c}{Plug-in} & \multicolumn{1}{c}{50} & \multicolumn{1}{c}{0.0000} & \multicolumn{1}{c}{0.0492} & \multicolumn{1}{c}{0.2476} & \multicolumn{1}{c}{74.5848} & \multicolumn{1}{c}{0.6757} & \multicolumn{1}{c}{73,222.6480} & \multicolumn{1}{c}{73,222.6480} \\
    \multicolumn{1}{c}{MELO} & \multicolumn{1}{c}{50} & \multicolumn{1}{c}{0.0000} & \multicolumn{1}{c}{0.0536} & \multicolumn{1}{c}{0.2125} & \multicolumn{1}{c}{0.5777} & \multicolumn{1}{c}{0.5394} & \multicolumn{1}{c}{78.8489} & \multicolumn{1}{c}{78.8489} \\
    \multicolumn{1}{c}{Plug-in} & \multicolumn{1}{c}{500} & \multicolumn{1}{c}{0.0000} & \multicolumn{1}{c}{0.0043} & \multicolumn{1}{c}{0.0223} & \multicolumn{1}{c}{0.0599} & \multicolumn{1}{c}{0.0706} & \multicolumn{1}{c}{1.4684} & \multicolumn{1}{c}{1.4684} \\
    \multicolumn{1}{c}{MELO} & \multicolumn{1}{c}{500} & \multicolumn{1}{c}{0.0000} & \multicolumn{1}{c}{0.0049} & \multicolumn{1}{c}{0.0216} & \multicolumn{1}{c}{0.0556} & \multicolumn{1}{c}{0.0663} & \multicolumn{1}{c}{1.2967} & \multicolumn{1}{c}{1.2967} \\
    \multicolumn{1}{c}{Plug-in} & \multicolumn{1}{c}{1,000} & \multicolumn{1}{c}{0.0000} & \multicolumn{1}{c}{0.0029} & \multicolumn{1}{c}{0.0134} & \multicolumn{1}{c}{0.0304} & \multicolumn{1}{c}{0.0355} & \multicolumn{1}{c}{0.5353} & \multicolumn{1}{c}{0.5353} \\
    \multicolumn{1}{c}{MELO} & \multicolumn{1}{c}{1,000} & \multicolumn{1}{c}{0.0000} & \multicolumn{1}{c}{0.0034} & \multicolumn{1}{c}{0.0136} & \multicolumn{1}{c}{0.0291} & \multicolumn{1}{c}{0.0355} & \multicolumn{1}{c}{0.4880} & \multicolumn{1}{c}{0.4880} \\
    \hline
    \hline
    \multicolumn{9}{c}{Mean Absolute Error} \\
    \hline
    \hline
    \multicolumn{1}{c}{Method} & \multicolumn{1}{c}{Sample size} & \multicolumn{1}{c}{Min} & \multicolumn{1}{c}{1st Qu.} & \multicolumn{1}{c}{Median} & \multicolumn{1}{c}{Mean} & \multicolumn{1}{c}{3rd Qu.} & \multicolumn{1}{c}{Max} & \multicolumn{1}{c}{Range} \\
    \hline
    \multicolumn{1}{c}{Plug-in*} & \multicolumn{1}{c}{20} & \multicolumn{1}{c}{0.0039} & \multicolumn{1}{c}{0.4984} & \multicolumn{1}{c}{0.8846} & \multicolumn{1}{c}{3.6623E+07} & \multicolumn{1}{c}{1.3727} & \multicolumn{1}{c}{1.3166E+10} & \multicolumn{1}{c}{1.3166E+10} \\
    \multicolumn{1}{c}{MELO**} & \multicolumn{1}{c}{20} & \multicolumn{1}{c}{0.0007} & \multicolumn{1}{c}{0.4102} & \multicolumn{1}{c}{0.7721} & \multicolumn{1}{c}{0.7900} & \multicolumn{1}{c}{1.0496} & \multicolumn{1}{c}{5.4523} & \multicolumn{1}{c}{5.4516} \\
    \multicolumn{1}{c}{Plug-in} & \multicolumn{1}{c}{50} & \multicolumn{1}{c}{0.0005} & \multicolumn{1}{c}{0.2218} & \multicolumn{1}{c}{0.4976} & \multicolumn{1}{c}{1.0006} & \multicolumn{1}{c}{0.8220} & \multicolumn{1}{c}{270.5968} & \multicolumn{1}{c}{270.5963} \\
    \multicolumn{1}{c}{MELO} & \multicolumn{1}{c}{50} & \multicolumn{1}{c}{0.0011} & \multicolumn{1}{c}{0.2315} & \multicolumn{1}{c}{0.4610} & \multicolumn{1}{c}{0.5525} & \multicolumn{1}{c}{0.7344} & \multicolumn{1}{c}{8.8797} & \multicolumn{1}{c}{8.8786} \\
    \multicolumn{1}{c}{Plug-in} & \multicolumn{1}{c}{500} & \multicolumn{1}{c}{0.0004} & \multicolumn{1}{c}{0.0658} & \multicolumn{1}{c}{0.1493} & \multicolumn{1}{c}{0.1870} & \multicolumn{1}{c}{0.2658} & \multicolumn{1}{c}{1.2118} & \multicolumn{1}{c}{1.2114} \\
    \multicolumn{1}{c}{MELO} & \multicolumn{1}{c}{500} & \multicolumn{1}{c}{0.0000} & \multicolumn{1}{c}{0.0697} & \multicolumn{1}{c}{0.1471} & \multicolumn{1}{c}{0.1822} & \multicolumn{1}{c}{0.2575} & \multicolumn{1}{c}{1.1387} & \multicolumn{1}{c}{1.1387} \\
    \multicolumn{1}{c}{Plug-in} & \multicolumn{1}{c}{1,000} & \multicolumn{1}{c}{0.0001} & \multicolumn{1}{c}{0.0540} & \multicolumn{1}{c}{0.1157} & \multicolumn{1}{c}{0.1370} & \multicolumn{1}{c}{0.1884} & \multicolumn{1}{c}{0.7316} & \multicolumn{1}{c}{0.7315} \\
    \multicolumn{1}{c}{MELO} & \multicolumn{1}{c}{1,000} & \multicolumn{1}{c}{0.0000} & \multicolumn{1}{c}{0.0579} & \multicolumn{1}{c}{0.1165} & \multicolumn{1}{c}{0.1346} & \multicolumn{1}{c}{0.1883} & \multicolumn{1}{c}{0.6985} & \multicolumn{1}{c}{0.6985} \\
    \hline
    \hline
    \multicolumn{9}{c}{Mean Absolute Percentage Error} \\
    \hline
    \hline
    \multicolumn{1}{c}{Method} & \multicolumn{1}{c}{Sample size} & \multicolumn{1}{c}{Min} & \multicolumn{1}{c}{1st Qu.} & \multicolumn{1}{c}{Median} & \multicolumn{1}{c}{Mean} & \multicolumn{1}{c}{3rd Qu.} & \multicolumn{1}{c}{Max} & \multicolumn{1}{c}{Range} \\
    \hline
    \multicolumn{1}{c}{Plug-in*} & \multicolumn{1}{c}{20} & \multicolumn{1}{c}{0.0033} & \multicolumn{1}{c}{0.4249} & \multicolumn{1}{c}{0.7541} & \multicolumn{1}{c}{3.1219E+07} & \multicolumn{1}{c}{1.1702} & \multicolumn{1}{c}{1.1223E+10} & \multicolumn{1}{c}{1.1223E+10} \\
    \multicolumn{1}{c}{MELO**} & \multicolumn{1}{c}{20} & \multicolumn{1}{c}{0.0006} & \multicolumn{1}{c}{0.3497} & \multicolumn{1}{c}{0.6582} & \multicolumn{1}{c}{0.6734} & \multicolumn{1}{c}{0.8947} & \multicolumn{1}{c}{4.6478} & \multicolumn{1}{c}{4.6471} \\
    \multicolumn{1}{c}{Plug-in} & \multicolumn{1}{c}{50} & \multicolumn{1}{c}{0.0004} & \multicolumn{1}{c}{0.1891} & \multicolumn{1}{c}{0.4241} & \multicolumn{1}{c}{0.8530} & \multicolumn{1}{c}{0.7007} & \multicolumn{1}{c}{230.6682} & \multicolumn{1}{c}{230.6677} \\
    \multicolumn{1}{c}{MELO} & \multicolumn{1}{c}{50} & \multicolumn{1}{c}{0.0009} & \multicolumn{1}{c}{0.1974} & \multicolumn{1}{c}{0.3929} & \multicolumn{1}{c}{0.4710} & \multicolumn{1}{c}{0.6261} & \multicolumn{1}{c}{7.5694} & \multicolumn{1}{c}{7.5685} \\
    \multicolumn{1}{c}{Plug-in} & \multicolumn{1}{c}{500} & \multicolumn{1}{c}{0.0003} & \multicolumn{1}{c}{0.0561} & \multicolumn{1}{c}{0.1273} & \multicolumn{1}{c}{0.1594} & \multicolumn{1}{c}{0.2266} & \multicolumn{1}{c}{1.0330} & \multicolumn{1}{c}{1.0326} \\
    \multicolumn{1}{c}{MELO} & \multicolumn{1}{c}{500} & \multicolumn{1}{c}{0.0000} & \multicolumn{1}{c}{0.0594} & \multicolumn{1}{c}{0.1254} & \multicolumn{1}{c}{0.1553} & \multicolumn{1}{c}{0.2195} & \multicolumn{1}{c}{0.9707} & \multicolumn{1}{c}{0.9707} \\
    \multicolumn{1}{c}{Plug-in} & \multicolumn{1}{c}{1,000} & \multicolumn{1}{c}{0.0001} & \multicolumn{1}{c}{0.0461} & \multicolumn{1}{c}{0.0986} & \multicolumn{1}{c}{0.1168} & \multicolumn{1}{c}{0.1606} & \multicolumn{1}{c}{0.6237} & \multicolumn{1}{c}{0.6236} \\
    \multicolumn{1}{c}{MELO} & \multicolumn{1}{c}{1,000} & \multicolumn{1}{c}{0.0000} & \multicolumn{1}{c}{0.0493} & \multicolumn{1}{c}{0.0993} & \multicolumn{1}{c}{0.1147} & \multicolumn{1}{c}{0.1606} & \multicolumn{1}{c}{0.5955} & \multicolumn{1}{c}{0.5954} \\
    \hline
    \hline
    \multicolumn{9}{l}{* We discard the ``$\infty$" values     ** We discard the ``NA" values and when $Plug-in$ takes ``$\infty$" values} \\
    \end{tabular}%
		}
  \label{Table2}%
\end{sidewaystable}

%% file: Table3.tex
\begin{sidewaystable}
  \centering
  \caption{Odds ratio problem: Mean Errors for $x=(1,0,0)$}
	{\scriptsize
        \begin{tabular}{rrrrrrrrr}
    \hline
    \hline
    \multicolumn{9}{c}{Mean Square Error} \\
    \hline
    \hline
    \multicolumn{1}{c}{Method} & \multicolumn{1}{c}{Sample size} & \multicolumn{1}{c}{Min} & \multicolumn{1}{c}{1st Qu.} & \multicolumn{1}{c}{Median} & \multicolumn{1}{c}{Mean} & \multicolumn{1}{c}{3rd Qu.} & \multicolumn{1}{c}{Max} & \multicolumn{1}{c}{Range} \\
    \hline
    \multicolumn{1}{c}{Plug-in*} & \multicolumn{1}{c}{20} & \multicolumn{1}{c}{0.0000} & \multicolumn{1}{c}{0.2868} & \multicolumn{1}{c}{1.2486} & \multicolumn{1}{c}{1.2183E+15} & \multicolumn{1}{c}{5.0878} & \multicolumn{1}{c}{2.8697E+17} & \multicolumn{1}{c}{2.8697E+17} \\
    \multicolumn{1}{c}{MELO**} & \multicolumn{1}{c}{20} & \multicolumn{1}{c}{0.0000} & \multicolumn{1}{c}{0.1521} & \multicolumn{1}{c}{1.0307} & \multicolumn{1}{c}{2.8656} & \multicolumn{1}{c}{2.8027} & \multicolumn{1}{c}{112.7688} & \multicolumn{1}{c}{112.7688} \\
    \multicolumn{1}{c}{Plug-in} & \multicolumn{1}{c}{50} & \multicolumn{1}{c}{0.0000} & \multicolumn{1}{c}{0.1201} & \multicolumn{1}{c}{0.4335} & \multicolumn{1}{c}{5.9867E+06} & \multicolumn{1}{c}{1.3330} & \multicolumn{1}{c}{5.6088E+09} & \multicolumn{1}{c}{5.6088E+009} \\
    \multicolumn{1}{c}{MELO} & \multicolumn{1}{c}{50} & \multicolumn{1}{c}{0.0000} & \multicolumn{1}{c}{0.1052} & \multicolumn{1}{c}{0.3735} & \multicolumn{1}{c}{3.0819} & \multicolumn{1}{c}{1.1375} & \multicolumn{1}{c}{471.5700} & \multicolumn{1}{c}{471.5700} \\
    \multicolumn{1}{c}{Plug-in} & \multicolumn{1}{c}{500} & \multicolumn{1}{c}{0.0000} & \multicolumn{1}{c}{0.0068} & \multicolumn{1}{c}{0.0374} & \multicolumn{1}{c}{0.0977} & \multicolumn{1}{c}{0.1151} & \multicolumn{1}{c}{2.7585} & \multicolumn{1}{c}{2.7585} \\
    \multicolumn{1}{c}{MELO} & \multicolumn{1}{c}{500} & \multicolumn{1}{c}{0.0000} & \multicolumn{1}{c}{0.0069} & \multicolumn{1}{c}{0.0377} & \multicolumn{1}{c}{0.0933} & \multicolumn{1}{c}{0.1120} & \multicolumn{1}{c}{2.5165} & \multicolumn{1}{c}{2.5165} \\
    \multicolumn{1}{c}{Plug-in} & \multicolumn{1}{c}{1,000} & \multicolumn{1}{c}{0.0000} & \multicolumn{1}{c}{0.0048} & \multicolumn{1}{c}{0.0205} & \multicolumn{1}{c}{0.0439} & \multicolumn{1}{c}{0.0524} & \multicolumn{1}{c}{0.5349} & \multicolumn{1}{c}{0.5349} \\
    \multicolumn{1}{c}{MELO} & \multicolumn{1}{c}{1,000} & \multicolumn{1}{c}{0.0000} & \multicolumn{1}{c}{0.0047} & \multicolumn{1}{c}{0.0202} & \multicolumn{1}{c}{0.0429} & \multicolumn{1}{c}{0.0538} & \multicolumn{1}{c}{0.5100} & \multicolumn{1}{c}{0.5100} \\
    \hline
    \hline
    \multicolumn{9}{c}{Mean Absolute Error} \\
    \hline
    \hline
    \multicolumn{1}{c}{Method} & \multicolumn{1}{c}{Sample size} & \multicolumn{1}{c}{Min} & \multicolumn{1}{c}{1st Qu.} & \multicolumn{1}{c}{Median} & \multicolumn{1}{c}{Mean} & \multicolumn{1}{c}{3rd Qu.} & \multicolumn{1}{c}{Max} & \multicolumn{1}{c}{Range}\\
    \hline
    \multicolumn{1}{c}{Plug-in*} & \multicolumn{1}{c}{20} & \multicolumn{1}{c}{0.0018} & \multicolumn{1}{c}{0.5355} & \multicolumn{1}{c}{1.1174} & \multicolumn{1}{c}{2.6221E+06} & \multicolumn{1}{c}{2.2556} & \multicolumn{1}{c}{5.3569E+08} & \multicolumn{1}{c}{5.3569E+08} \\
    \multicolumn{1}{c}{MELO**} & \multicolumn{1}{c}{20} & \multicolumn{1}{c}{0.0052} & \multicolumn{1}{c}{0.3900} & \multicolumn{1}{c}{1.0152} & \multicolumn{1}{c}{1.2250} & \multicolumn{1}{c}{1.6741} & \multicolumn{1}{c}{10.6193} & \multicolumn{1}{c}{10.6141} \\
    \multicolumn{1}{c}{Plug-in} & \multicolumn{1}{c}{50} & \multicolumn{1}{c}{0.0008} & \multicolumn{1}{c}{0.3465} & \multicolumn{1}{c}{0.6584} & \multicolumn{1}{c}{100.5325} & \multicolumn{1}{c}{1.1546} & \multicolumn{1}{c}{74,892.1046} & \multicolumn{1}{c}{74,892.1038} \\
    \multicolumn{1}{c}{MELO} & \multicolumn{1}{c}{50} & \multicolumn{1}{c}{0.0004} & \multicolumn{1}{c}{0.3243} & \multicolumn{1}{c}{0.6112} & \multicolumn{1}{c}{0.9356} & \multicolumn{1}{c}{1.0665} & \multicolumn{1}{c}{21.7157} & \multicolumn{1}{c}{21.7153} \\
    \multicolumn{1}{c}{Plug-in} & \multicolumn{1}{c}{500} & \multicolumn{1}{c}{0.0003} & \multicolumn{1}{c}{0.0823} & \multicolumn{1}{c}{0.1934} & \multicolumn{1}{c}{0.2381} & \multicolumn{1}{c}{0.3392} & \multicolumn{1}{c}{1.6609} & \multicolumn{1}{c}{1.6606} \\
    \multicolumn{1}{c}{MELO} & \multicolumn{1}{c}{500} & \multicolumn{1}{c}{0.0003} & \multicolumn{1}{c}{0.0833} & \multicolumn{1}{c}{0.1942} & \multicolumn{1}{c}{0.2343} & \multicolumn{1}{c}{0.3346} & \multicolumn{1}{c}{1.5863} & \multicolumn{1}{c}{1.5860} \\
    \multicolumn{1}{c}{Plug-in} & \multicolumn{1}{c}{1,000} & \multicolumn{1}{c}{0.0005} & \multicolumn{1}{c}{0.0693} & \multicolumn{1}{c}{0.1432} & \multicolumn{1}{c}{0.1671} & \multicolumn{1}{c}{0.2290} & \multicolumn{1}{c}{0.7313} & \multicolumn{1}{c}{0.7309} \\
    \multicolumn{1}{c}{MELO} & \multicolumn{1}{c}{1,000} & \multicolumn{1}{c}{0.0003} & \multicolumn{1}{c}{0.0687} & \multicolumn{1}{c}{0.1420} & \multicolumn{1}{c}{0.1657} & \multicolumn{1}{c}{0.2321} & \multicolumn{1}{c}{0.7141} & \multicolumn{1}{c}{0.7138} \\
    \hline
    \hline
    \multicolumn{9}{c}{Mean Absolute Percentage Error } \\
    \hline
    \hline
    \multicolumn{1}{c}{Method} & \multicolumn{1}{c}{Sample size} & \multicolumn{1}{c}{Min} & \multicolumn{1}{c}{1st Qu.} & \multicolumn{1}{c}{Median} & \multicolumn{1}{c}{Mean} & \multicolumn{1}{c}{3rd Qu.} & \multicolumn{1}{c}{Max} & \multicolumn{1}{c}{Range} \\
    \hline
    \multicolumn{1}{c}{Plug-in*} & \multicolumn{1}{c}{20} & \multicolumn{1}{c}{0.0008} & \multicolumn{1}{c}{0.2389} & \multicolumn{1}{c}{0.4986} & \multicolumn{1}{c}{1.1700E+06} & \multicolumn{1}{c}{1.0065} & \multicolumn{1}{c}{2.3903E+08} & \multicolumn{1}{c}{2.3903E+08} \\
    \multicolumn{1}{c}{MELO**} & \multicolumn{1}{c}{20} & \multicolumn{1}{c}{0.0023} & \multicolumn{1}{c}{0.1740} & \multicolumn{1}{c}{0.4530} & \multicolumn{1}{c}{0.5466} & \multicolumn{1}{c}{0.7470} & \multicolumn{1}{c}{4.7384} & \multicolumn{1}{c}{4.7361} \\
    \multicolumn{1}{c}{Plug-in} & \multicolumn{1}{c}{50} & \multicolumn{1}{c}{0.0003} & \multicolumn{1}{c}{0.1546} & \multicolumn{1}{c}{0.2938} & \multicolumn{1}{c}{44.8586} & \multicolumn{1}{c}{0.5152} & \multicolumn{1}{c}{33,417.6096} & \multicolumn{1}{c}{33,417.6092} \\
    \multicolumn{1}{c}{MELO} & \multicolumn{1}{c}{50} & \multicolumn{1}{c}{0.0002} & \multicolumn{1}{c}{0.1447} & \multicolumn{1}{c}{0.2727} & \multicolumn{1}{c}{0.4175} & \multicolumn{1}{c}{0.4759} & \multicolumn{1}{c}{9.6897} & \multicolumn{1}{c}{9.6896} \\
    \multicolumn{1}{c}{Plug-in} & \multicolumn{1}{c}{500} & \multicolumn{1}{c}{0.0001} & \multicolumn{1}{c}{0.0367} & \multicolumn{1}{c}{0.0863} & \multicolumn{1}{c}{0.1063} & \multicolumn{1}{c}{0.1514} & \multicolumn{1}{c}{0.7411} & \multicolumn{1}{c}{0.7410} \\
    \multicolumn{1}{c}{MELO} & \multicolumn{1}{c}{500} & \multicolumn{1}{c}{0.0001} & \multicolumn{1}{c}{0.0372} & \multicolumn{1}{c}{0.0867} & \multicolumn{1}{c}{0.1046} & \multicolumn{1}{c}{0.1493} & \multicolumn{1}{c}{0.7078} & \multicolumn{1}{c}{0.7077} \\
    \multicolumn{1}{c}{Plug-in} & \multicolumn{1}{c}{1,000} & \multicolumn{1}{c}{0.0002} & \multicolumn{1}{c}{0.0309} & \multicolumn{1}{c}{0.0639} & \multicolumn{1}{c}{0.0746} & \multicolumn{1}{c}{0.1022} & \multicolumn{1}{c}{0.3263} & \multicolumn{1}{c}{0.3261} \\
    \multicolumn{1}{c}{MELO} & \multicolumn{1}{c}{1,000} & \multicolumn{1}{c}{0.0002} & \multicolumn{1}{c}{0.0307} & \multicolumn{1}{c}{0.0634} & \multicolumn{1}{c}{0.0739} & \multicolumn{1}{c}{0.1035} & \multicolumn{1}{c}{0.3187} & \multicolumn{1}{c}{0.3185} \\
    \hline
    \hline
    \multicolumn{9}{l}{* We discard the ``$\infty$" values     **We discard the ``NA" values and when $Plug-in$ takes ``$\infty$" values} \\
    \end{tabular}%
		}
  \label{Table3}%
\end{sidewaystable}

%% file: Table4.tex
\begin{sidewaystable}
	\centering
	\caption{Demand and supply model: Mean Errors.}
	{\tiny{
	\begin{tabular}{cccccccccccccccccc}
		\hline
& & & \multicolumn{4}{c}{Mean Squared Error} & \multicolumn{4}{c}{Mean Absolute Error} &  \multicolumn{4}{c}{Mean Absolute Percentage Error}\\
\hline
				
						Signal/Noise	&	Method	&	Sample size	&	$\beta_1$	&	$\beta_2$	&	$\alpha_1$	&	$\alpha_2$	&	$\beta_1$	&	$\beta_2$	&	$\alpha_1$	&	$\alpha_2$	&	$\beta_1$	&	$\beta_2$	&	$\alpha_1$	&	$\alpha_2$	\\
						\hline																											
						\multirow{15}{*}{0.1}	&	2SLS	&	\multirow{3}{*}{20}	&	884.83	&	5,030.75	&	2,433,378.77	&	14,396,948.28	&	5.35	&	10.76	&	54.55	&	129.84	&	669.36\%	&	717.06\%	&	4545.69\%	&	12983.79\%	\\
						&	Analytical MELO	&		&	0.80	&	5.70	&	1.56	&	6.05	&	0.80	&	1.91	&	1.17	&	1.95	&	99.56\%	&	127.20\%	&	97.44\%	&	194.72\%	\\
						&	Computational MELO	&		&	0.81	&	5.75	&	1.57	&	6.10	&	0.80	&	1.91	&	1.17	&	1.95	&	99.80\%	&	127.64\%	&	97.45\%	&	195.45\%	\\
						&	2SLS	&	\multirow{3}{*}{50}	&	638.92	&	544.20	&	14,826.33	&	47,907.91	&	5.07	&	5.07	&	9.17	&	13.27	&	633.62\%	&	337.88\%	&	763.99\%	&	1326.56\%	\\
						&	Analytical MELO	&		&	0.78	&	2.48	&	1.42	&	2.96	&	0.79	&	1.25	&	1.08	&	1.36	&	98.17\%	&	83.41\%	&	89.60\%	&	136.29\%	\\
						&	Computational MELO	&		&	0.78	&	2.50	&	1.42	&	2.98	&	0.79	&	1.25	&	1.07	&	1.36	&	98.26\%	&	83.58\%	&	89.57\%	&	136.49\%	\\
						&	2SLS	&	\multirow{3}{*}{100}	&	429.43	&	633.86	&	201.59	&	145.63	&	4.87	&	5.07	&	4.17	&	3.63	&	609.32\%	&	338.20\%	&	347.14\%	&	362.62\%	\\
						&	Analytical MELO	&		&	0.75	&	1.55	&	1.22	&	1.66	&	0.76	&	1.00	&	0.98	&	1.02	&	95.43\%	&	66.35\%	&	81.74\%	&	102.41\%	\\
						&	Computational MELO	&		&	0.75	&	1.55	&	1.23	&	1.66	&	0.76	&	1.00	&	0.98	&	1.03	&	95.49\%	&	66.40\%	&	81.70\%	&	102.51\%	\\
						&	2SLS	&	\multirow{3}{*}{1000}	&	107.71	&	84.95	&	3.44	&	1.20	&	2.20	&	1.78	&	0.74	&	0.53	&	274.66\%	&	118.61\%	&	61.67\%	&	52.89\%	\\
						&	Analytical MELO	&		&	0.31	&	0.31	&	0.25	&	0.22	&	0.44	&	0.44	&	0.40	&	0.37	&	55.02\%	&	29.51\%	&	33.17\%	&	37.17\%	\\
						&	Computational MELO	&		&	0.31	&	0.31	&	0.25	&	0.22	&	0.44	&	0.44	&	0.40	&	0.37	&	55.01\%	&	29.50\%	&	33.19\%	&	37.18\%	\\
						&	2SLS	&	\multirow{3}{*}{20000}	&	0.03	&	0.03	&	0.02	&	0.02	&	0.14	&	0.12	&	0.11	&	0.10	&	17.72\%	&	8.33\%	&	9.02\%	&	10.40\%	\\
						&	Analytical MELO	&		&	0.03	&	0.02	&	0.02	&	0.02	&	0.14	&	0.12	&	0.11	&	0.10	&	17.24\%	&	8.16\%	&	8.89\%	&	10.34\%	\\
						&	Computational MELO	&		&	0.03	&	0.02	&	0.02	&	0.02	&	0.14	&	0.12	&	0.11	&	0.10	&	17.24\%	&	8.16\%	&	8.89\%	&	10.34\%	\\
						\hline																													
						\multirow{15}{*}{0.5}	&	2SLS	&	\multirow{3}{*}{20}	&	889.82	&	864.22	&	731.46	&	140.04	&	3.43	&	2.74	&	3.44	&	1.77	&	429.33\%	&	182.80\%	&	286.66\%	&	177.36\%	\\
						&	Analytical MELO	&		&	0.51	&	0.46	&	0.45	&	0.45	&	0.59	&	0.54	&	0.54	&	0.53	&	73.95\%	&	36.26\%	&	44.64\%	&	53.17\%	\\
						&	Computational MELO	&		&	0.52	&	0.46	&	0.45	&	0.46	&	0.59	&	0.55	&	0.53	&	0.54	&	74.20\%	&	36.42\%	&	44.57\%	&	53.54\%	\\
						&	2SLS	&	\multirow{3}{*}{50}	&	336.38	&	98.10	&	8.28	&	2.69	&	1.98	&	1.27	&	0.66	&	0.50	&	247.52\%	&	84.92\%	&	54.86\%	&	49.68\%	\\
						&	Analytical MELO	&		&	0.32	&	0.25	&	0.22	&	0.22	&	0.45	&	0.40	&	0.37	&	0.37	&	56.46\%	&	26.55\%	&	30.76\%	&	36.71\%	\\
						&	Computational MELO	&		&	0.32	&	0.25	&	0.23	&	0.22	&	0.45	&	0.40	&	0.37	&	0.37	&	56.60\%	&	26.61\%	&	30.87\%	&	36.79\%	\\
						&	2SLS	&	\multirow{3}{*}{100}	&	5.94	&	2.47	&	0.26	&	0.16	&	0.68	&	0.52	&	0.36	&	0.30	&	85.36\%	&	34.90\%	&	30.27\%	&	30.22\%	\\
						&	Analytical MELO	&		&	0.19	&	0.15	&	0.15	&	0.12	&	0.35	&	0.31	&	0.30	&	0.27	&	43.14\%	&	20.54\%	&	25.36\%	&	27.49\%	\\
						&	Computational MELO	&		&	0.19	&	0.15	&	0.15	&	0.12	&	0.35	&	0.31	&	0.30	&	0.28	&	43.23\%	&	20.56\%	&	25.39\%	&	27.51\%	\\
						&	2SLS	&	\multirow{3}{*}{1000}	&	0.03	&	0.02	&	0.02	&	0.01	&	0.12	&	0.11	&	0.10	&	0.08	&	15.62\%	&	7.47\%	&	8.28\%	&	8.23\%	\\
						&	Analytical MELO	&		&	0.02	&	0.02	&	0.02	&	0.01	&	0.12	&	0.11	&	0.10	&	0.08	&	15.22\%	&	7.32\%	&	8.23\%	&	8.19\%	\\
						&	Computational MELO	&		&	0.02	&	0.02	&	0.02	&	0.01	&	0.12	&	0.11	&	0.10	&	0.08	&	15.22\%	&	7.32\%	&	8.23\%	&	8.19\%	\\
						&	2SLS	&	\multirow{3}{*}{20000}	&	0.00	&	0.00	&	0.00	&	0.00	&	0.03	&	0.02	&	0.02	&	0.02	&	3.47\%	&	1.64\%	&	1.80\%	&	2.07\%	\\
						&	Analytical MELO	&		&	0.00	&	0.00	&	0.00	&	0.00	&	0.03	&	0.02	&	0.02	&	0.02	&	3.46\%	&	1.63\%	&	1.79\%	&	2.07\%	\\
						&	Computational MELO	&		&	0.00	&	0.00	&	0.00	&	0.00	&	0.03	&	0.02	&	0.02	&	0.02	&	3.46\%	&	1.63\%	&	1.79\%	&	2.08\%	\\
						\hline																													
						\multirow{15}{*}{1}	&	2SLS	&	\multirow{3}{*}{20}	&	47.94	&	13.73	&	0.39	&	0.21	&	1.01	&	0.71	&	0.42	&	0.33	&	125.70\%	&	47.61\%	&	34.82\%	&	33.21\%	\\
						&	Analytical MELO	&		&	0.22	&	0.17	&	0.18	&	0.15	&	0.37	&	0.33	&	0.33	&	0.30	&	45.99\%	&	21.75\%	&	27.55\%	&	29.93\%	\\
						&	Computational MELO	&		&	0.22	&	0.18	&	0.19	&	0.15	&	0.37	&	0.33	&	0.33	&	0.30	&	46.44\%	&	21.88\%	&	27.80\%	&	30.06\%	\\
						&	2SLS	&	\multirow{3}{*}{50}	&	0.60	&	0.29	&	0.09	&	0.06	&	0.37	&	0.29	&	0.23	&	0.20	&	46.14\%	&	19.17\%	&	18.87\%	&	19.63\%	\\
						&	Analytical MELO	&		&	0.13	&	0.08	&	0.08	&	0.06	&	0.28	&	0.23	&	0.21	&	0.19	&	35.07\%	&	15.39\%	&	17.64\%	&	19.09\%	\\
						&	Computational MELO	&		&	0.13	&	0.08	&	0.08	&	0.06	&	0.28	&	0.23	&	0.21	&	0.19	&	35.16\%	&	15.42\%	&	17.65\%	&	19.10\%	\\
						&	2SLS	&	\multirow{3}{*}{100}	&	0.08	&	0.05	&	0.04	&	0.03	&	0.22	&	0.18	&	0.17	&	0.14	&	27.46\%	&	12.19\%	&	13.76\%	&	14.13\%	\\
						&	Analytical MELO	&		&	0.07	&	0.05	&	0.04	&	0.03	&	0.20	&	0.17	&	0.16	&	0.14	&	25.51\%	&	11.47\%	&	13.28\%	&	13.87\%	\\
						&	Computational MELO	&		&	0.07	&	0.05	&	0.04	&	0.03	&	0.20	&	0.17	&	0.16	&	0.14	&	25.53\%	&	11.47\%	&	13.28\%	&	13.88\%	\\
						&	2SLS	&	\multirow{3}{*}{1000}	&	0.01	&	0.00	&	0.00	&	0.00	&	0.06	&	0.05	&	0.05	&	0.04	&	7.66\%	&	3.67\%	&	4.13\%	&	4.11\%	\\
						&	Analytical MELO	&		&	0.01	&	0.00	&	0.00	&	0.00	&	0.06	&	0.05	&	0.05	&	0.04	&	7.60\%	&	3.64\%	&	4.12\%	&	4.11\%	\\
						&	Computational MELO	&		&	0.01	&	0.00	&	0.00	&	0.00	&	0.06	&	0.05	&	0.05	&	0.04	&	7.60\%	&	3.64\%	&	4.12\%	&	4.11\%	\\
						&	2SLS	&	\multirow{3}{*}{20000}	&	0.00	&	0.00	&	0.00	&	0.00	&	0.01	&	0.01	&	0.01	&	0.01	&	1.73\%	&	0.82\%	&	0.90\%	&	1.04\%	\\
						&	Analytical MELO	&		&	0.00	&	0.00	&	0.00	&	0.00	&	0.01	&	0.01	&	0.01	&	0.01	&	1.73\%	&	0.82\%	&	0.90\%	&	1.04\%	\\
						&	Computational MELO	&		&	0.00	&	0.00	&	0.00	&	0.00	&	0.01	&	0.01	&	0.01	&	0.01	&	1.73\%	&	0.82\%	&	0.90\%	&	1.04\%	\\
						\hline																													
						\multirow{15}{*}{5}	&	2SLS	&	\multirow{3}{*}{20}	&	0.01	&	0.01	&	0.01	&	0.01	&	0.09	&	0.08	&	0.07	&	0.06	&	11.57\%	&	5.17\%	&	5.97\%	&	6.01\%	\\
						&	Analytical MELO	&		&	0.01	&	0.01	&	0.01	&	0.01	&	0.09	&	0.08	&	0.07	&	0.06	&	11.42\%	&	5.11\%	&	5.93\%	&	5.99\%	\\
						&	Computational MELO	&		&	0.01	&	0.01	&	0.01	&	0.01	&	0.09	&	0.08	&	0.07	&	0.06	&	11.42\%	&	5.11\%	&	5.94\%	&	5.99\%	\\
						&	2SLS	&	\multirow{3}{*}{50}	&	0.01	&	0.00	&	0.00	&	0.00	&	0.06	&	0.05	&	0.04	&	0.04	&	7.62\%	&	3.30\%	&	3.57\%	&	3.81\%	\\
						&	Analytical MELO	&		&	0.01	&	0.00	&	0.00	&	0.00	&	0.06	&	0.05	&	0.04	&	0.04	&	7.58\%	&	3.29\%	&	3.55\%	&	3.80\%	\\
						&	Computational MELO	&		&	0.01	&	0.00	&	0.00	&	0.00	&	0.06	&	0.05	&	0.04	&	0.04	&	7.57\%	&	3.29\%	&	3.55\%	&	3.80\%	\\
						&	2SLS	&	\multirow{3}{*}{100}	&	0.00	&	0.00	&	0.00	&	0.00	&	0.04	&	0.04	&	0.03	&	0.03	&	5.19\%	&	2.34\%	&	2.68\%	&	2.77\%	\\
						&	Analytical MELO	&		&	0.00	&	0.00	&	0.00	&	0.00	&	0.04	&	0.03	&	0.03	&	0.03	&	5.18\%	&	2.33\%	&	2.67\%	&	2.77\%	\\
						&	Computational MELO	&		&	0.00	&	0.00	&	0.00	&	0.00	&	0.04	&	0.03	&	0.03	&	0.03	&	5.18\%	&	2.33\%	&	2.67\%	&	2.77\%	\\
						&	2SLS	&	\multirow{3}{*}{1000}	&	0.00	&	0.00	&	0.00	&	0.00	&	0.01	&	0.01	&	0.01	&	0.01	&	1.52\%	&	0.73\%	&	0.83\%	&	0.82\%	\\
						&	Analytical MELO	&		&	0.00	&	0.00	&	0.00	&	0.00	&	0.01	&	0.01	&	0.01	&	0.01	&	1.52\%	&	0.73\%	&	0.83\%	&	0.82\%	\\
						&	Computational MELO	&		&	0.00	&	0.00	&	0.00	&	0.00	&	0.01	&	0.01	&	0.01	&	0.01	&	1.52\%	&	0.73\%	&	0.83\%	&	0.82\%	\\
						&	2SLS	&	\multirow{3}{*}{20000}	&	0.00	&	0.00	&	0.00	&	0.00	&	0.00	&	0.00	&	0.00	&	0.00	&	0.35\%	&	0.16\%	&	0.18\%	&	0.21\%	\\
						&	Analytical MELO	&		&	0.00	&	0.00	&	0.00	&	0.00	&	0.00	&	0.00	&	0.00	&	0.00	&	0.35\%	&	0.16\%	&	0.18\%	&	0.21\%	\\
						&	Computational MELO	&		&	0.00	&	0.00	&	0.00	&	0.00	&	0.00	&	0.00	&	0.00	&	0.00	&	0.35\%	&	0.16\%	&	0.18\%	&	0.21\%	\\
		\hline
		\hline
	\end{tabular}}}
	\label{Table4}
\end{sidewaystable}